\documentstyle[psfig,aps]{revtex}
\tightenlines
\def\Journal#1#2#3#4{{#1} {\bf #2}, #3 (#4)}

\def\AP{{\em Ann. Phys.}}
\def\APG{{\em Ann. Physik}}
\def\ADP{{\em Adv. in Phys.}}
\def\CMP{{\em Comm. Math. Phys.}}
\def\IJMPA{{\em Int. J. of Mod. Phys.}}

\def\JETP{{\em Sov. Phys. J. JETP}}
\def\JETPL{{\em JETP Lett.}}

\def\JPF{{\em J. Phys.} F}

\def\PHM{{\em Phil. Mag.}}

\def\PRL{\em Phys. Rev. Lett.}
\def\PR{{\em Phys. Rev.}}

\def\PRB{{\em Phys. Rev.} B}
\def\PRD{{\em Phys. Rev.} D}

\def\PREPC{{\em Phys. Rep.} C}

\def\RMP{{\em Rev. Mod. Phys.}}

\def\RPP{{\em Rep. Prog. Phys.}}
\def\ZPB{{\em Z. Phys.} B}

\newcommand{\be}{\begin{equation}}
\newcommand{\ee}{\end{equation}}
\newcommand{\bea}{\begin{eqnarray}}
\newcommand{\eea}{\end{eqnarray}}
\newcommand{\hf}{{1\over2}}
\newcommand{\nonu}{\nonumber\\}
\def\eq#1{(\ref{#1})}
\def\la{\langle}
\def\ra{\rangle}
\def\cd#1{{\cal D}[#1]}
\def\tr{{\mathrm Tr}}
\def\ord#1{{\cal O}(#1)}
\def\mb#1{{\mathbf#1}}
\def\mr#1{{\mathrm#1}}
\def\tD{\tilde D}
\def\tDi{{\tilde D}^{-1}}
\def\ve{V^{\mathrm ext}}
\def\ae{A^{\mathrm ext}}
\def\oper{{\cal O}}
\def\psid{\psi^\dagger}
\def\pp{{\cal P}}

\def\br{\hskip -0.2cm /}
\def\sign{{\mathrm{sign}}}

\begin{document}
\title{Current-density functional for disordered systems}
\author{Janos Polonyi\thanks{polonyi@fresnel.u-strasbg.fr}}
\address{Laboratory of Theoretical Physics, Louis Pasteur University,
Strasbourg, France}
\address{Department of Atomic Physics, L. E\"otv\"os University,
Budapest, Hungary}
\date{\today}
\maketitle
\begin{abstract}
The effective action for the current and density is shown to
satisfy an evolution equation, the functional generalization of 
Callan-Symanzik equation. The solution describes the dependence
of the one-particle irreducible vertex functions on the strength
of the quenched disorder and the annealed Coulomb interaction. 
The result is non-perturbative, no small parameter is assumed. 
The a.c. conductivity is obtained by 
the numerical solution of the evolution equation on finite lattices 
in the absence of the Coulomb interaction. The static limit
is performed and the conductivity is found to be vanishing beyond 
a certain threshold of the impurity strength.
\end{abstract}

\section{Introduction}
The currently used method to obtain observables for disordered systems
is to perform a quenched averaging, an averaging of the Green
functions over random external field \cite{edw}. The computational
algorithm boils down either to the analytical continuation
in the number of replicas \cite{replica}, or to the introduction
of fictious particles related to the real ones by super-transformations
\cite{susy} or to the use of the Keldysh contour in computing
loop-integrals \cite{keldysh}. The impact of weak disorder on the
conductivity has nicely been captured by resumming the dominant
graphs \cite{langer} in a self-consistent manner \cite{ww}. But
the whole scheme is based on perturbation expansion and it remains
unclear how to treat strongly disordered systems and understand
the transition to localization. This feature appears
to be a serious limiting factor in the description of strongly
correlated systems, as well, where the interactions are supposed
to be non-perturbative.
The goal of this paper is to demonstrate that methods, well established
in other domains of physics, in particular the density functional
scheme and the functional extension of the renormalization group
might be useful to tackle this problem. Namely, a new method is
proposed for the systemical, non-perturbative computation of the
queneched average of the one-particle irreducible (1PI) vertex
functions.

The density functional \cite{dft} turned out to be a powerful formalism
in handle the ground states. Its virtue is that the variational
setting gives a large degrees of flexibility for improvements of the
density functional. The extension to include the
current in the variational techniques has already been suggested, as well
\cite{cft}. This method is improved on the following points.
First, the density-current functional is generalized for time-dependent
density and current configurations in order to encompass the dynamics
and to arrive at a closed set of equations to determine this functional.
Second, a constructive definition of the functional is used
which is based on the Legendre transform of the generator functional for the
connected Green functions for the density and current. In fact, it has
been realized lately that such an effective action can be identified with
the density functional \cite{system,kornel}. Third, the functional
generalization \cite{frg,pozs} of the traditional renormalization group
procedure \cite{trg} is used in the internal space \cite{int} to
obtain the effective action. The evolution equation describes the
changes generated by the gradual increase of the amplitude of the
fluctuations.
The only approximation committed in this scheme is the truncation of the
effective action, the projection of the functional renormalization
group equation into a constrained functional space. One hopes to
carry out this truncation in a manner which is independent of the amplitude
of the fluctuations and the resulting scheme will be non-perturbative.
We shall apply gradient expansion and retain arbitrarily high orders
which seems to be a consistent scheme in determing the kinetic transport
coefficients.

The functional renormalization group method has already been applied
for disordered systems, namely to the pinning of an elastic 
system in a random potential \cite{pin}. A random external potential
introduced in an elastic system generates infinitely many relevant or 
marginal coupling constants and their treatment required the functional 
extension of the renormalization group procedure. The U.V. cut-off 
was gradually decreased and the resulting renormalization group equation was
solved within the framework of the replica method. In the present work we 
do not touch the cut-off, instead we follow the evolution of the system when
a coupling constant is gradually 'turned on'. Such a modification
of the rules of the renormalization group method allows us to
embark directly the issue of quenched averaging and offers an
alternative to the replica method.

The initial condition is usually chosen to be a theory with such
parameters which allow an approximative determination of the effective
action. The evolution equation generates the 'renormalized trajectory'
along which the physics changes since the quantum/thermal fluctuations
are turned on gradually. The integration of the
evolution equation from the initial condition with small fluctuations
to the strongly coupled theory under consideration represents a general
purpose algorithm to solve strongly interactive systems.

It is shown below that the quenched average of the generator functional
for the connected Green functions for the density satisfies a simple
evolution equation when the strength of the disorder is gradually
changed. This is not surprising for weak disorder since
the perturbative averaging over the impurity potential introduces correlations
which are represented by introducing 'interaction vertices' for the density
and the corresponding perturbation series involves Green functions with 
increasing number of density insertions. The evolution equation is a
differential equation for the generator functional for the
Green functions which expresses the change induced by the infinitesimal
increase of the strength of the disorder in terms of functional derivatives
with respect to the external source coupled to the density.
In case of an interactive system the coupling constants of the annealed 
interactions are evolved, too. The equation is obtained
without any reference to small parameters in the dynamics and
is valid non-perturbatively. The evolution equation, being a
functional differential equation represents a wonderful mathematical problem.
In lacking theorems and other general support one has to truncate
this equation by projecting it into a restricted functional space.
The ordinary functions characterizing the functionals in this space,
the running coupling strengths in the language of the renormalization
group, will then be evolved by differential equations. It is rather
natural that local functionals are easier to handle in a truncation
scheme. Therefore the evolution equation is recasted in terms of
a local functional, the effective action for the density.

The generator functionals are well suited to the linear response
formalism since the introduction of the external sources can easily
be achieved by functional derivation with respect to appropriately
chosen external sources. Therefore we include the electric current
besides the density in our formalism in order to keep track of
issues related to charge transport and will use the generator
functional for the connected Green functions for the current and density
operators and their effective action. The formal evolution equation will be
presented for the case of electrons with Coulomb interaction but the
numerical solution will be discussed for the non-interactive case only.

We start in section \ref{genfuns} with the introduction of the model and its
generator functionals. The average of the generator functional
for the connected Green functions for the current and the density operators
over the impurity field is defined in the next step. Its functional Legendre
transform, the effective action or the current-density functional is the main
object of this paper. Connection to the density functional theory is briefly 
commented, as well. It is furthermore mentioned that the generator functional is well suited
to the linear response computations except that the causal propagators
have to be replaced by the retarded or advanced ones in order to implement
the physical boundary conditions in time. This gives a simple short-cut
to re-derive the Kubo expressions for the transport coefficients.

The evolution equation method is discussed in section \ref{evoes}. Though the
complete interactive case is covered, the equation is studied in a more
detailed manner for non-interactive particles only. The functional differential
equation is approximated by truncating either the powers of the fields
or the power of the space-time gradients acting on the fields. The
former is useful to establish a contact with the resummation methods
of the perturbation series, the latter is more appropriate for the
computation of the transport coefficients.

Before solving the evolution equation we have to construct the initial
condition. This is the subject of section \ref{initcos}.
The initial condition is defined in the absence of disorder where the
effective action is obtained by retaining infinitely many orders
of the gradient expansion. First we compute the
one-loop two-point functions for the current and density and
construct the generator functional. After that we perform
the functional Legendre transformation to find the effective action.

The brief survey of the numerical solution of the evolution equation is
presented in section \ref{numers}. The extrapolation of the conductivity
to the homogeneous case found out to be rather difficult for
weak disorder and should be studied in a more detailed manner. 
But strong disorder suppresses the long range modes and the
finite size dependence and allows to study the conductivity on lattices
up to the size $320^3$. The preliminary results
presented in this paper show a discontinuity developing in the
static limit, in particular the d.c. conductivity drops to zero
at a certain strength of the disorder. It is natural to
identify this point with the onset of localization.

Section \ref{concl} is for the conclusion. The technical details of the functional
calculus, a formal relation between the evolution equation and the replica method
and the perturbative evaluation of the two-point functions for the
current and density are presented in the appendices together with the
derivation of the Ward identities which are needed to assure the 
finiteness of the d.c. conductivity.

\section{Current-density functional}\label{genfuns}
The model will be introduced in this section together with the
generator functional for its connected Green functions, its quenched
average and functional Legendre transform, the effective action.

\subsection{The model}
We consider spinless electrons propagating on an external electromagnetic
potentials $\ve_x$ and $\ae_x$ and eventually interacting with the
Coulomb potential. The path integral
representation of the vacuum-to-vacuum amplitude is
\be
Z=\int\cd{\psi}\cd{\psid}\cd{u}e^{{i\over\hbar}S[\psid,\psi,u]},
\ee
with
\be
S[\psid,\psi,u]=\int_x\left\{\psid_x
\left[i\hbar\partial_t-{1\over2m}\left({\hbar\over i}\partial+\ae_x\right)^2
+\mu+e\ve_x+eu_x\right]\psi_x-\hf(\partial u_x)^2\right\},
\ee
c.f. Appendix \ref{notations} for the notations. We shall use a
space-time lattice with spacings $a_j=a$, $j=1,2,3$ and $a_0=a_\tau$.
There several reasons to introduce lattice regularization.
The most important one is that the non-relativistic models 
are usually non-renormalizable and need cut-off for their definition.
But even if we are satisfied by considering non-interacting electrons,
as will be the case in most of this paper, which are U.V. finite
the functions appearing in the description such as the Green functions develop
I.R. singularities. This requires
a special care in the numerical treatment of the one-loop
integral of the evolution equation. The breakdown of this
loop integral into a large but finite sum can be done in a singularity free
manner only if the gauge invariance properties are not spoiled in the 
finite sum. The only known method to achieve this is to use lattice 
regularization. The regulated action will be
\be\label{action}
S[\psid,\psi,u]=a^6a^2_\tau\sum_{x',x}\psid_{x'}G^{-1}_{x',x}\psi_x
+{a^3a_\tau\over2}\sum_x(\nabla u_x)^2,
\ee
in terms of the inverse propagator
\bea\label{prop}
G^{-1}_{x',x}[\ve,\ae]
&=&-{\hbar\over a_\tau}\left(\delta^K_{x',x}
-\delta^K_{x',x+\hat0}U_0(t,\mb{x})\right)\nonu
&&+{\hbar^2\over2ma^2}\sum_{j>0}\left(\delta^K_{x'+\hat j,x}U^*_j(x-\hat j)
+\delta^K_{x'-\hat j,x}U_j(x)-2\delta^K_{x',x}\right),
\eea
where $\hat\mu=a_\mu e_\mu$ and the link variables
\be
U_0(x)=e^{-{i\over\hbar}a_\tau(\mu+e\ve_x+eu_x)},~~~~
U_j(x)=e^{-{i\over\hbar}ea\ae_{j,x}}
\ee
are used to assure gauge invariance \cite{lgt}.

The generator functional for the connected Green functions of local
operators $\oper_{\alpha,x}$,$\tilde\oper_{\tilde\alpha,x}$
coupled to the external sources $\sigma_{\alpha,x}$,
$\tilde\sigma_{\tilde\alpha,x}$, $\alpha=0,1,2,3$, $\tilde\alpha=1,2,3$, is
\be\label{genfunction}
e^{{i\over\hbar}W_0[\sigma]}=\int\cd{u}\cd{\psi}\cd{\psid}
e^{{i\over\hbar}\psid\cdot(G^{-1}[\ve,\ae]+\sigma\br)\cdot\psi
+{i\over2\hbar}\partial u\cdot\partial u},
\ee
where $\psid\cdot\sigma\br\cdot\psi=\sigma_0\cdot\oper_0+\sigma_j\cdot\oper_j
+\tilde\sigma_j\cdot\tilde\oper_j$. We shall choose
\bea\label{choice}
\psid\cdot\oper_{0,x}\cdot\psi&=&{\delta S\over\delta eV^\mr{ext}_x}
=i\psid_{x+\hat0}U_0(x)\psi_x,\nonu
\psid\cdot\oper_{j,x}\cdot\psi
&=&{\delta S\over\delta eA^\mr{ext}_{j,x}}
={i\hbar\over2ma}\sum_{j>0}\left[\psid_{x+\hat j}U_j(x)\psi_x
-\psid_xU^*_j(x)\psi_{x+\hat j}\right],\nonu
\delta_{j,k}\delta_{x,y}\psid\cdot\tilde\oper_{j,x}\cdot\psi
&=&m{\delta\psid\cdot\oper_{j,x}\cdot\psi\over\delta eA^\mr{ext}_{k,y}}
=\hf\delta_{j,k}\delta_{x,y}\left[\psid_{x+\hat j}U_j(x)\psi_x
+\psid_xU^*_j(x)\psi_{x+\hat j}\right].
\eea
The operators $\tilde\oper_{j,x}$ are introduced in order to derive
simplify the formal steps in deriving the Ward identities.

\subsection{Averaging over the disorder}\label{avdis}
Let us suppose that the presence of the impurities can be described by a
static potential $v_\mb{x}$, with Gaussian distribution \cite{edw}. The
quenched average of observables corresponding to a typical static impurity
configuration can be constructed by means of the generator functional
for averaged connected Green functions,
\be\label{impdist}
W[\sigma]={\int\cd{v}e^{-{1\over2g}\int_\mb{x}v^2_\mb{x}}
W_0[\sigma_0+v,\sigma_1,\ldots]
\over\int\cd{v}e^{-{1\over2g}\int_\mb{x}v^2_\mb{x}}}.
\ee
Two remarks are in order at this point. First, local impurity potential
distribution is assumed for the sake of simplicity
only, each step below can be reproduced with any other Gaussian
distribution. Second, such a simple averaging is available for
the Green functions of the density only, the operator which couples linearly
to the impurity potential.

Since the impurity potential appears as an additional external source
coupled to the density one expects that its modification of the dynamics
can be obtained by means of the external potential $\sigma_0$.
A simple explicit equation which realizes this possibility can immediately
be obtained by rewriting \eq{impdist} as
\be
W[\sigma]=W_0\left[\sigma_0+{\delta\over\delta j},\sigma_1,\ldots\right]
e^{{g\over2}\int_\mb{x}j^2_\mb{x}}_{\ \ \ \ \ \ \vert j=0}.
\ee
This equation can be used to generate the usual perturbation series \cite{edw}.

Another variant of Eq. \eq{impdist} which will be useful for the
derivation of the evolution equation is established by using the 'equation
of motion' for the impurity potential. The equation of motion
expresses the invariance of the path integral with respect to
the infinitesimal shift of the integral variable,
\be\label{eqmo}
0=\int\cd{\phi}e^{{i\over\hbar}S[\phi]}{\delta S[\phi]\over\delta\phi_x}.
\ee
The finite shift $v_\mb{p}\to v_\mb{p}-i\sigma_{\omega=0,\mb{p}}/T$
where where $T$ is the Euclidean time extent of the system
carried out for the Fourier transform gives
\be\label{quav}
W[\sigma]={\int\cd{v}e^{-{1\over2g}\int_\mb{p}
(v_\mb{-p}-{1\over T}\sigma_{0,\omega=0,\mb{-p}})
(v_\mb{p}-{1\over T}\sigma_{0,\omega=0,\mb{p}})}
W^v_0[\sigma]\over\int\cd{v}e^{-{1\over2g}\int_\mb{x}v^2_\mb{x}}}.
\ee
where shifted functional is
\be
W^v_0[\sigma_0,\sigma_1,\ldots]
=W_0[\sigma_{0,\omega=0}=Tv,\sigma_{0,\omega\not=0},\sigma_1,\ldots].
\ee
This equation explains the use of the internal space blocking method
for the computation of quenched averages. In this blocking procedure
a parameter which controls the amplitude of the fluctuations, say the
mass for relativistic scalar theories, is driving the evolution. In
the present case $g$, the average amplitude square of the impurity field,
plays a similar role. In fact, the impurity field fluctuations just smear out
the generator functional within an interval of the size $\ord{\sqrt{g}}$
in the dependence on the external source. The evolution equation
introduced below follows as the increase of $g$ gradually 'opens up' a
large interval for the source in which the disorder acts.

The functional $W[\sigma]$ will be useful to express the change of the
dynamics under an infinitesimal change of a parameter of the dynamics
but leads to complications when such an evolution equation
is truncated. The complications arise from the fact that $W[\sigma]$
belongs to the class of non-local functionals which are very hard to
characterize and handle.
The way out form this problem is the introduction of the
current-density functional, the effective action for the composite
operators $\oper_\alpha$ as the Legendre transform of $W[\sigma]$,
\be\label{legtro}
\Gamma[\rho]=-W[\sigma]+\sigma_\alpha\cdot\rho_\alpha
+\tilde\sigma_{\tilde\alpha}\cdot\rho_{\tilde\alpha},~~~~
\rho_{\alpha,x}={\delta W[\sigma]\over\delta\sigma_{\alpha,x}},~~~~
\tilde\rho_{\tilde\alpha,x}
={\delta W[\sigma]\over\delta\tilde\sigma_{\tilde\alpha,x}}.
\ee
This functional is local and offers useful truncation schemes.

A relation which will later be used a number of occasions is
\be\label{wg}
\int_{x'}{\delta^2W[\sigma]\over\delta\sigma_{\alpha,x}
\delta\sigma_{\gamma,x'}}
{\delta^2\Gamma[\rho]\over\delta\rho_{\gamma,x'}\delta\rho_{\beta,x''}}
=\delta_{x,x''}\delta_{\alpha,\beta}.
\ee

\subsection{Variational method}
The Hohenberg-Kohn theorems \cite{dft} assure that
(i) the density functional $\Gamma[\rho]$ which gives the ground
state energy in the sector of the Fock space belonging to a given
charge density $\rho_x$ is a well defined functional and
(ii) the minimum of the density functional is reached at the
density profile of the true ground state where its value reproduces
the energy of the ground state. We generalize the density functional
first to include the current $\Gamma[\rho]\to\Gamma[\rho_\mu]$, where
$\rho_0$ and $\rho_j$ ($j=1,2,3$) is the expectation value
of the density and the current, respectively. After that we allow
the density and the current to become space and time dependent.

This generalization is realized by the effective action \eq{legtro}.
In fact, the inverse transformation
is carried out by the change of variable
\be
\sigma_{\alpha,x}={\delta\Gamma[\rho]\over\delta\rho_{\alpha,x}},~~~
\tilde\sigma_{\tilde\alpha,x}
={\delta\Gamma[\rho]\over\delta\tilde\rho_{\tilde\alpha,x}},
\ee
and this relation indicates that the ground state corresponds to the
stationary points of the effective action. The transformation
\eq{legtro} can be used to argue that the effective action
gives the quantum action ($-i\hbar$ times the logarithm of the
transition amplitude) for real time or the free energy for
imaginary time. Therefore the Euclidean effective action is
bounded from below and the ground state is found at its
minimum. Therefore the effective action plays the role of the
current-density functional whose minimization leads us to the ground state.

\subsection{Linear response}\label{linresp}
As mentioned in the Introduction, the generator functional is
well suited to the linear response formalism. To demonstrate this
point let us consider the system in the presence of an additional
external electromagnetic field $A_\mu$ imposed for $t<0$ by making
the replacement $\ve_t\to V_t'=\ve_t+A_{0,t}$ and $\ae_t\to A_t'=\ae_t+A_t$ in
\eq{prop} where the eventual time dependence arising from
the external fields is shown explicitly.
For the computation of the conductivity $j_k=\sigma_{k,\ell}E_\ell$,
we use the electric field $\mb{E}=-\partial_t\mb{A}$ and
the interaction Hamiltonian $H'=-e\int_xA_{k,x}j_{k,x}$.
The ground state $|0_t\ra$, following the external fields
in an adiabatic manner might show time dependence, as well.
The expectation value $\la0_t|j_{k,t,\mb{x}}|0_t\ra$ of the induced gauge
invariant current for $t=0$ is
\bea
\la0_0|j_{k,0,\mb{x}}|0_0\ra&=&{i\hbar e\over2ma}\la0_0|
\psid_{s,0,\mb{x}+k}U'_k(x)\psi_{s,0,\mb{x}}
-\psid_{s,0,\mb{x}}U^{'*}_k(x)\psi_{s,0,\mb{x}+k}|0_0\ra\nonu
&\approx& e\la0_0|\oper_{k,0,\mb{x}}|0_0\ra
-{e^2\over2m}A_{k,0,\mb{x}}\la0_0|\psid_{s,0,\mb{x}+k}U_k(x)\psi_{s,0,\mb{x}}
+\psid_{s,0,\mb{x}}U^*_k(x)\psi_{s,0,\mb{x}+k}|0_0\ra\nonu
&=&e\int_y\Theta(-y^0){\delta\la0_0|\oper_{k,0,\mb{x}}|0_0\ra
\over\delta A_{\ell,y}}A_{\ell,y}-{e^2\over m}A_{k,0,\mb{x}}\tilde\rho^*.
\eea
Since the bra $\la0_0|$ is the time reversed version of the ket $|0_0\ra$
in the absence of spin the functional derivative in the last equation
contains the contributions of the bra and the ket with a relative
negative sign which corresponds to the commutator between the 
external source and the observable
in the Kubo formula. The time inversion can formally be
realized in the framework of the path integral formalism by
using external field which is symmetrized with respect to
the time inversion and by taking the complex conjugate of
the integrand for $t>0$. This latter step brings the
product of the Heavyside function and the functional
derivative into the retarded current-current Green function,
\be
\la0_0|j_{k,0,\mb{x}}|0_0\ra
=-e^2\int_y\tD^{Rk,\ell}_{(0,\mb{x}),y}A_{\ell,y}
-{e^2\over m}A_{k,0,\mb{x}}\tilde\rho^*,
\ee
and we have
\be
\sigma_{k,\ell}={ie^2\over\omega}\left({\tilde\rho^*\over m}\delta^{k,\ell}
+\tD^{Rk,\ell}_{\omega,\mb{q}}\right)
={ie^2\over\omega}\left({\tilde\rho^*\over m}\delta^{k,\ell}
+Im\tD^{k,\ell}_{\omega+i0^+,\mb{q}}\right)
\ee
in agreement with the Kubo formula. In the last equation we have
the imaginary part of the causal propagator because the
current operator is hermitian.

\section{Evolution equation}\label{evoes}
We introduce the evolution equation and discuss its structure for
non-interactive systems first. The generalization which includes
interaction, the Coulomb potential in particular is presented briefly
later.

In the traditional renormalization group method one orders
the degrees of freedom $\phi_1,\phi_2,\ldots,\phi_N$
according to their momenta, $p_1<p_2<\ldots<p_N$
and eliminates them one-by one, starting with the perturbative,
high momentum modes. The corresponding blocking for the action is
\be
e^{iS_{N-1}(\phi_1,\phi_2,\ldots,\phi_{N-1})}
=\int d\phi_Ne^{iS_N(\phi_1,\phi_2,\ldots,\phi_N)}.
\ee
Such a method may be called external space renormalization group
since the momentum is a scale in the external space, i.e. in the space-time.
As an alternative, one may order the fluctuations
according to their amplitude, the scale parameter in the internal
space of quantum field theory \cite{int}. The evolution from
$\lambda=0$ to $\lambda=\infty$ of the partition function
\be
Z_\lambda=\int_{-\lambda}^\lambda d\phi_1\cdots
\int_{-\lambda}^\lambda d\phi_Ne^{iS_N(\phi_1,\phi_2,\ldots,\phi_N)}
\ee
takes into account the fluctuations in the order of increasing
amplitude. A more practical, smooth cut-off in the internal space
can be implemented by introducing a suppression term in the action
$S[\phi]\to S[\phi]+S_\lambda[\phi]$, such as
\be
S_\lambda[\phi]=-{\lambda\over2}\phi\cdot\phi.
\ee
In relativistic field theory this term generates mass and the
realizes the Callan-Symanzik scheme \cite{calsym}.

The evolution equation can easiest be derived by starting with the
rather trivial identity
\be
{d\over d\lambda}Z[j]={d\over d\lambda}\int\cd{\phi}
e^{{i\over\hbar}S[\phi]+{i\over\hbar}\phi\cdot j}
={i\over\hbar}\int\cd{\phi}e^{{i\over\hbar}S[\phi]+{i\over\hbar}\phi\cdot j}
{dS[\phi]\over d\lambda}.
\ee
In the second step the expectation value standing in the right
hand side is expressed by acting with the functional derivatives
with respect to the source $j$ on the generator functional $Z[j]$
and the result is a functional differential equation for $Z[j]$.
The harmless looking first step is actually a generalization
of the Hellman-Feynman theorem \cite{helfe} for the time
evolution operator. Different choices for $\lambda$ yield important,
well known equations. The choice of $\lambda$ as a field
variable $\phi(x)$ leads to the Schwinger-Dyson equation.
When $\lambda$ is chosen to be
a parameter of a transformation of the field variable which
leaves a part or the full action invariant then the resulting
equations are the Ward identities.

The disadvantage
of the internal space renormalization group method is that the
renormalized trajectory does not belong a to fixed physics anymore
and the evolution of the control parameter $\lambda$ has little to do
with the scale dependence in the theory. But the advantage is
the flexibility in choosing $\lambda$.
In fact, by identifying $\lambda$ with the Planck constant or
a gauge coupling one can construct evolution schemes where
the semiclassical structure remains invariant or the gauge
invariance is preserved explicitely, respectively. The solution
of the evolution from a perturbative initial condition with
small amplitude fluctuations to the desired level of fluctuations
provides us a systematical non-perturbative algorithm to
compute irreducible Green functions.

\subsection{Non-interacting particles}
It was mentioned in Section \ref{avdis} that the effects of the disorder
and the external source $\sigma_0$ are related. We follow up this
remark by showing that the infinitesimal change of the strength
of the disorder can be reproduced by acting on $W[\sigma]$ with functional
derivatives with respect to $\sigma_0$. In order to keep the measure of the
strength of the disorder, $g$, explicit in the equations we make the
replacement $g\to g\lambda$ and look for the evolution equation in $\lambda$,
\bea
\dot\Gamma[\rho]&=&-\dot W[\sigma]\nonu
&=&-{\int\cd{v}e^{-{1\over2g\lambda}
\int_\mb{p}|v_\mb{-p}-{1\over T}\sigma_{0,\omega=0,\mb{-p}}|^2}
W^v_0[\sigma]\int_\mb{p}|v_\mb{-p}-{1\over T}\sigma_{0,\omega=0,\mb{-p}}|^2
\over2g\lambda^2\int\cd{v}e^{-{1\over2g\lambda}\int_\mb{x}v^2_\mb{x}}}\nonu
&&+{1\over2g\lambda^2}{\int\cd{v}e^{-{1\over2g\lambda}\int_\mb{p}
|v_\mb{-p}-{1\over T}\sigma_{0,\omega=0,\mb{-p}}|^2}
W^v_0[\sigma]\over\int\cd{v}e^{-{1\over2g\lambda}\int_\mb{x}v^2_\mb{x}}}
{\int\cd{v}e^{-{1\over2g\lambda}\int_\mb{x}v_\mb{x}^2}
\int_\mb{x}v_\mb{x}^2\over\int\cd{v}e^{-{1\over2g\lambda}
\int_\mb{x}v^2_\mb{x}}}\nonu
&=&-{g\over2}\int_\mb{p}{\delta^2W[\sigma]
\over\delta\sigma_{0,\omega=0,\mb{-p}}\delta\sigma_{0,\omega=0,\mb{p}}}\nonu
&=&-{g\over2}\int_\mb{p}
\left[{\delta^2\Gamma[\rho]\over\delta\rho\delta\rho}\right]
^{-1}_{(0,\omega=0,\mb{-p}),(0,\omega=0,\mb{p})},
\eea
where the dot stands for $\partial_\lambda$. We shall impose
the initial conditions at $\lambda=0$. It is shown in Appendix
\ref{repl} that this formalism is consistent with the replica
method so long the latter is applicable.

The functional differential equations are too difficult to handle
and the step of central importance is to find a truncation scheme
which renders this equation useful. We consider two truncations schemes
following the Landau-Ginzburg strategy of a double expansion in the
amplitude and the gradient of the fields. The truncation in the amplitude
has more pedagogical value, it provides a relation between
the resummation of the perturbation series and the evolution equation.
The truncation in the powers of the momenta, the gradient expansion,
seems to be a natural approximation scheme for the effective action since
the averaging over the impurity field leaves the ground state
translation invariant and the conductivity, as other transport parameters,
are closely related to the Taylor expansion of the 1PI Green functions
in the external momentum.

\subsection{Coulomb interaction}
We considered the system of non-interacting electrons so far.
But the evolution equation method can be used for interactive
systems, too. The non-trivial point is that both the annealed
and the quenched averages can be computed in a parallel manner
by generalizing the method of Ref. \cite{kornel} for disordered
systems.

The generator functional for the connected Green functions
for real time involving the rescaled photon field $u\to u/e$ is
\be
e^{{i\over\hbar}W_0[\sigma,j]}
=\int\cd{u}\cd{\psi}\cd{\psid}e^{{i\over\hbar}
\psid_s\cdot(G^{-1}+\sigma\br)\cdot\psi_s
+{i\over2\hbar e^2}\partial u\cdot\partial u
+{i\over\hbar}(j+\psid_s\psi_s)\cdot u}
\ee
We make the rescaling $e^2\to e^2\lambda$ and find the
evolution equation for the annealed average
\bea
\dot W_0[\sigma,j]&=&-{1\over2e^2\lambda^2}\int_x\Delta_x
\la u_xu_y\ra_{\vert x=y}\nonu
&=&{1\over2e^2\lambda^2}\left(i\hbar\tr\Delta
{\delta^2W_0[\sigma,j]\over\delta j\delta j}
-{\delta W_0[\sigma,j]\over\delta j}\cdot\Delta\cdot
{\delta W_0[\sigma,j]\over\delta j}\right),
\eea
in the electric charge. The functional we need is effective action
corresponding to the quenched average of the connected Green functions,
\be
\Gamma[\rho,w]=-W[\sigma,j]+\sigma_\alpha\cdot\rho_\alpha
+j\cdot w,~~~~
\rho_{\alpha,x}={\delta W[\sigma,j]\over\delta\sigma_{\alpha,x}},~~~~
w_x={\delta W[\sigma,j]\over\delta j_x},
\ee
where
\be
W[\sigma,j]={\int\cd{v}e^{-{1\over2g}\int_\mb{x}v^2_\mb{x}}
W_0[\sigma_0+v,\sigma_1,\ldots,j]
\over\int\cd{v}e^{-{1\over2g}\int_\mb{x}v^2_\mb{x}}}.
\ee

\begin{figure}
\centerline{\psfig{file=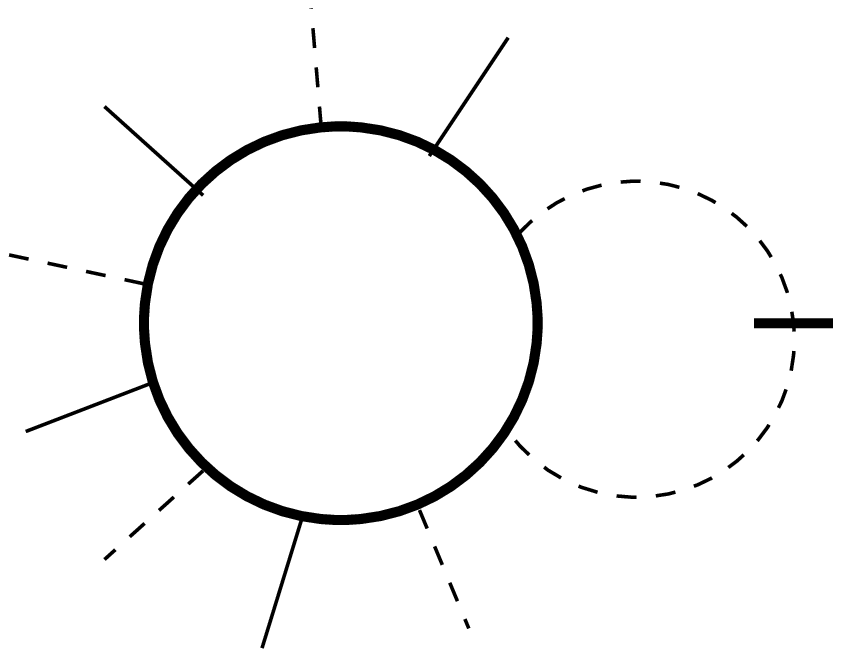,height=3.cm,width=4.cm,angle=0}}
\centerline{(a)}
\centerline{\psfig{file=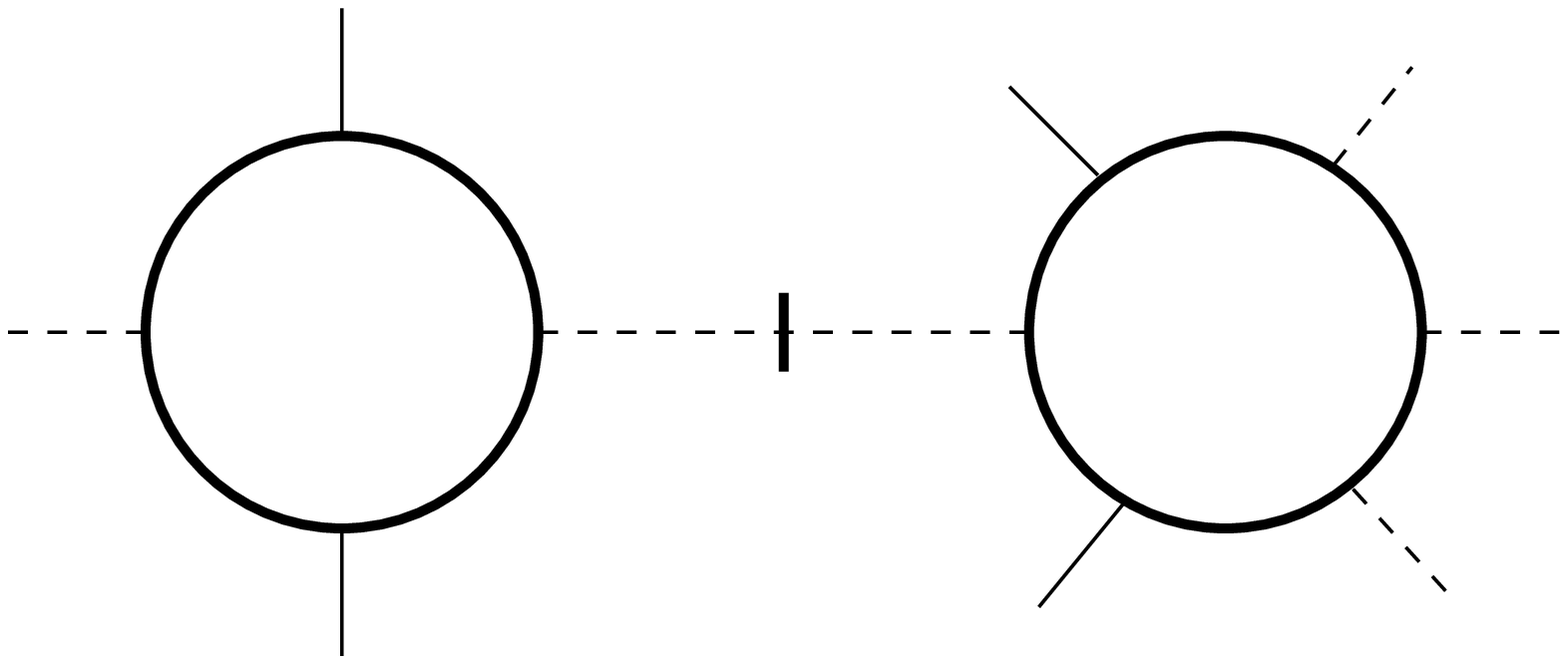,height=3.cm,width=8.cm,angle=0}}
\centerline{(b)}
\caption{Graphs contributing to the evolution equations of the
generator functional $W[\sigma,j]$ in $e^2$. The graph (a)
and (b) correspond to the one-particle irreducible and reducible
contributions, respectively. The solid and dashed line represent the
current-density and photon insertions, respectively
and and the circle with $n$ legs denotes the $n$-point
connected Green function. The cut stands for
the inverse photon propagator. \label{wev}}
\end{figure}

In order to perform the quenched averaging by evolving the system
we make the rescaling $g\to g\lambda$ in the distribution of the
impurity potential and find
\bea\label{evgamc}
\dot\Gamma[\rho,w]&=&-\dot W[\sigma,j]\nonu
&=&-{1\over2g\lambda^2}{\int\cd{v}e^{-{1\over2g}\int_\mb{p}
|v_\mb{-p}-{1\over T}\sigma_{0,\omega=0,\mb{-p}}|^2}
W^v_0[\sigma,j]\int_\mb{p}|v_\mb{-p}-{1\over T}\sigma_{0,\omega=0,\mb{-p}}|^2
\over\int\cd{v}e^{-{1\over2g}\int_\mb{x}v^2_\mb{x}}}\nonu
&&+{V\over2\lambda}{\int\cd{v}e^{-{1\over2g}\int_\mb{p}
|v_\mb{-p}-{1\over T}\sigma_{0,\omega=0,\mb{-p}}|^2}
W^v_0[\sigma,j]\over\int\cd{v}e^{-{1\over2g}\int_\mb{x}v^2_\mb{x}}}\nonu
&&+{1\over2e^2\lambda^2}{\int\cd{v}e^{-{1\over2g}\int_\mb{x}v^2_\mb{x}}
\left(i\hbar\tr\Delta{\delta^2W_0[\sigma_0+v,\sigma_1,\ldots,j]
\over\delta j\delta j}
-{\delta W_0[\sigma_0+v,\sigma_1,\ldots,j]\over\delta j}\cdot\Delta\cdot
{\delta W_0[\sigma_0+v,\sigma_1,\ldots,j]\over\delta j}\right)\over
\int\cd{v}e^{-{1\over2g}\int_\mb{x}v^2_\mb{x}}}.
\eea
The last line, the piece which generates the Coulomb interaction is
represented graphically in Fig. \ref{wev}.

\begin{figure}
\centerline{\psfig{file=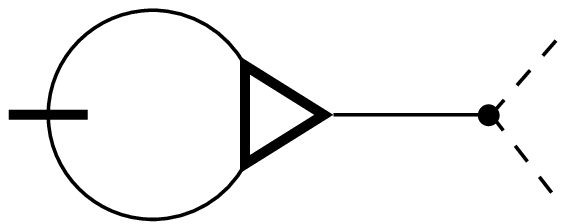,height=2.cm,width=5.cm,angle=0}}
\centerline{(a)}
\centerline{\psfig{file=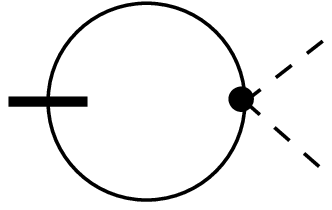,height=2.cm,width=3.5cm,angle=180}}
\centerline{(b)}
\caption{Graphs contributions to the evolution equations of
$L[\rho,w]$ in $\lambda$. The solid and dashed lines represent the propagators
$(\delta^2\Gamma/\delta\rho\delta\rho)^{-1}$ and
$(\delta^2\Gamma/\delta w\delta w)^{-1}$, respectively
and the triangular the vertex $\delta^3\Gamma/\delta\rho\delta\rho\delta\rho$.
The dot is for the functional derivatives $\delta L/\delta\rho$ and
$\delta^2L/\delta\rho\delta\rho$. The cut stands for
the correlation vertex $g$. \label{lev}}
\end{figure}

The formal difference between the annealed and the quenched averages appears
in the manner the disconnected parts are treated in the last two terms
on the right hand side of the second equation. In order to linearize
the right hand side in the functionals which facilitates the computation of
the quenched average we introduce a new functional
\be
K_0[\sigma,j]={\delta W_0[\sigma,j]\over\delta j}\cdot\Delta\cdot
{\delta W_0[\sigma,j]\over\delta j}
\ee
whose quenched average
\be
K[\sigma,j]={\int\cd{v}e^{-{1\over2g}\int_\mb{x}v^2_\mb{x}}
K_0[\sigma_0+v,\sigma_1,\ldots,j]
\over\int\cd{v}e^{-{1\over2g}\int_\mb{x}v^2_\mb{x}}}
\ee
satisfies the evolution equation
\bea
\dot K[\sigma,j]&=&-{1\over2g\lambda^2}
{\int\cd{v}e^{-{1\over2g}\int_\mb{p}
|v_\mb{-p}-{1\over T}\sigma_{0,\omega=0,\mb{-p}}|^2}K^v_0[\sigma,j]
\int_\mb{p}|v_\mb{-p}-{1\over T}\sigma_{0,\omega=0,\mb{-p}}|^2
\over\int\cd{v}e^{-{1\over2g\lambda}\int_xv^2_x}}\nonu
&&+{V\over2\lambda}{\int\cd{v}e^{-{1\over2g\lambda}\int_xv^2_x}
K_0[\sigma_0+v,\sigma_1,\ldots,j]
\over\int\cd{v}e^{-{1\over2g\lambda}\int_xv^2_x}}\nonu
&=&-{g\over2}\int_\mb{z}{\delta^2K[\sigma,j]\over
\delta\sigma_{0,0,\mb{z}}\delta\sigma_{0,0,\mb{z}}}.
\eea
We write $L[\rho,w]=K[\sigma,j]$ and the corresponding evolution equation is
\bea\label{kqev}
\dot L[\rho,w]&=&-{g\over2}\int_{\mb{z},x}
{\delta^2\rho_x\over\delta\sigma_{0,0,\mb{z}}\delta\sigma_{0,0,\mb{z}}}
{\delta L[\rho,w]\over\delta\rho_x}
-{g\over2}\int_{\mb{z},x,y}{\delta^2L[\rho,w]\over\delta\rho_x\delta\rho_y}
{\delta\rho_x\over\delta\sigma_{0,0,\mb{z}}}
{\delta\rho_y\over\delta\sigma_{0,0,\mb{z}}}\nonu
&=&-{g\over2}\int_{\mb{z},x}
{\delta^3W[\sigma,j]\over\delta\sigma_x\delta\sigma_{0,0,\mb{z}}\delta\sigma_{0,0,\mb{z}}}
{\delta L[\rho,w]\over\delta\rho_x}
-{g\over2}\int_{\mb{z},x,y}{\delta^2L[\rho,w]\over\delta\rho_x\delta\rho_y}
{\delta^2W[\sigma,j]\over\delta\sigma_x\delta\sigma_{0,0,\mb{z}}}
{\delta^2W[\sigma,j]\over\delta\sigma_y\delta\sigma_{0,0,\mb{z}}}\nonu
&=&-{g\over2}\int_{\mb{z},x}{\delta\over\delta\sigma_x}
\left({\delta^2\Gamma[\rho,w]\over\delta\rho\delta\rho}\right)^{-1}
_{(0,0,\mb{z}),(0,0,\mb{z})}
{\delta L[\rho,w]\over\delta\rho_x}\nonu
&&-{g\over2}\int_{\mb{z},x,y}{\delta^2L[\rho,w]\over\delta\rho_x\delta\rho_y}
\left({\delta^2\Gamma[\rho,w]\over\delta\rho\delta\rho}\right)^{-1}_{x,(0,0,\mb{z})}
\left({\delta^2\Gamma[\rho,w]\over\delta\rho\delta\rho}\right)^{-1}_{y,(0,0,\mb{z})}
\nonu
&=&-{g\over2}\int_{\mb{z},x,y,u}
\left({\delta^2\Gamma[\rho,w]\over\delta\rho\delta\rho}\right)^{-1}
_{(0,0,\mb{z}),y}{\delta\over\delta\sigma_x}
{\delta^2\Gamma[\rho,w]\over\delta\rho_y\delta\rho_u}
\left({\delta^2\Gamma[\rho,w]\over\delta\rho\delta\rho}\right)^{-1}_{u,(0,0,\mb{z})}
{\delta L[\rho,w]\over\delta\rho_x}\nonu
&&-{g\over2}\int_\mb{z}
\left[\left({\delta^2\Gamma[\rho,w]\over\delta\rho\delta\rho}\right)^{-1}
\cdot{\delta^2L[\rho,w]\over\delta\rho\delta\rho}\cdot
\left({\delta^2\Gamma[\rho,w]\over\delta\rho\delta\rho}\right)^{-1}
\right]_{(0,0,\mb{z}),(0,0,\mb{z})}\nonu
&=&-{g\over2}\int_\mb{z}
\left[\left({\delta^2\Gamma[\rho,w]\over\delta\rho\delta\rho}\right)^{-1}
\cdot\left(\int_x{\delta L[\rho,w]\over\delta\rho_x}{\delta\over\delta\sigma_x}
{\delta^2\Gamma[\rho,w]\over\delta\rho\delta\rho}
+{\delta^2L[\rho,w]\over\delta\rho\delta\rho}\right)\cdot
\left({\delta^2\Gamma[\rho,w]\over\delta\rho\delta\rho}\right)^{-1}
\right]_{(0,0,\mb{z}),(0,0,\mb{z})}\\
&=&-{g\over2}\int_\mb{z}
\left({\delta^2\Gamma[\rho,w]\over\delta\rho\delta\rho}\right)^{-1}
\cdot\left(\int_{x,y}{\delta L[\rho,w]\over\delta\rho_x}
\left({\delta^2\Gamma[\rho,w]\over\delta\rho\delta\rho}\right)^{-1}_{x,y}
{\delta^3\Gamma[\rho,w]\over\delta\rho\delta\rho\delta\rho_y}
+{\delta^2L[\rho,w]\over\delta\rho\delta\rho}\right)\cdot
\left({\delta^2\Gamma[\rho,w]\over\delta\rho\delta\rho}\right)^{-1}
_{(0,0,\mb{z}),(0,0,\mb{z})}\nonumber
\eea
which is shown graphically in Fig. \ref{lev}.

Therefore the \eq{evgamc} can be written as
\bea\label{gamtev}
\dot\Gamma[\rho,w]
&=&-{g\over2}\int_\mb{x}{\delta^2W[\sigma]\over\delta\sigma_{0,0,\mb{x}}}
+{1\over2e^2\lambda^2}\left\{i\hbar\tr\left[\Delta\left({\delta^2W[\sigma,j]
\over\delta j\delta j}-K[\sigma,j]\right)\right]\right\}\nonu
&=&-{g\over2}\int_\mb{z}\left[{\delta^2\Gamma[\rho,w]\over
\delta\rho\delta\rho}\right]^{-1}_{(0,0,\mb{z}),(0,0,\mb{z})}
+{1\over2e^2\lambda^2}\left\{i\hbar\tr\Delta\cdot\left[
\left({\delta^2\Gamma[\rho,w]\over\delta w\delta w}\right)^{-1}
-L[\rho,w]\right]\right\}.
\eea
The system of equations \eq{kqev} and \eq{gamtev} is closed
and describes the evolution of the interactive system.

\subsection{$\ord{\rho^4}$ truncation}\label{rhonegy}
\begin{figure}
\centerline{\psfig{file=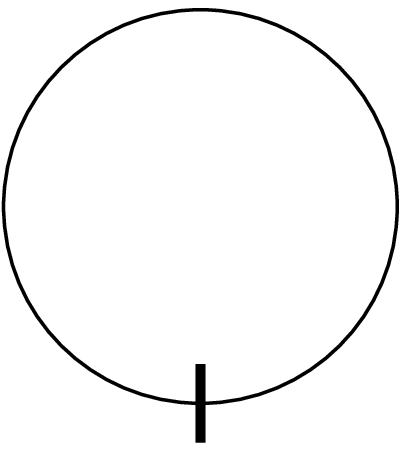,height=1.cm,width=1.cm,angle=0}}
\centerline{(a)}
\centerline{\psfig{file=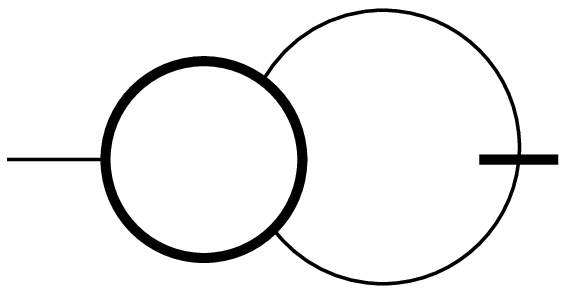,height=1.2cm,width=2.4cm,angle=0}}
\centerline{(b)}
\centerline{\psfig{file=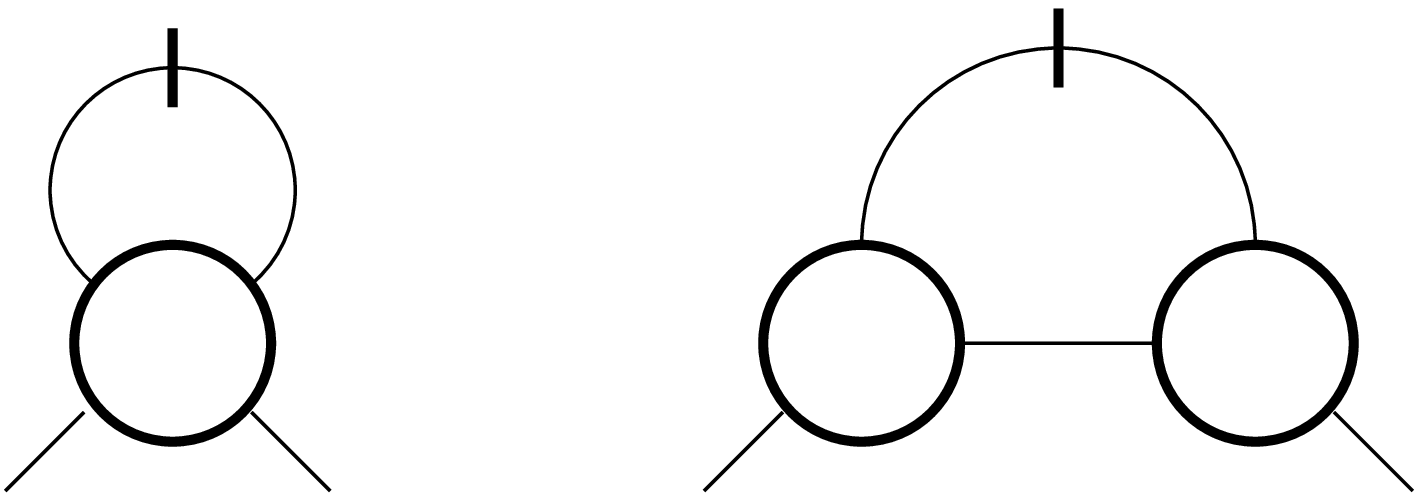,height=2.cm,width=6.cm,angle=0}}
\centerline{(c)}
\centerline{\psfig{file=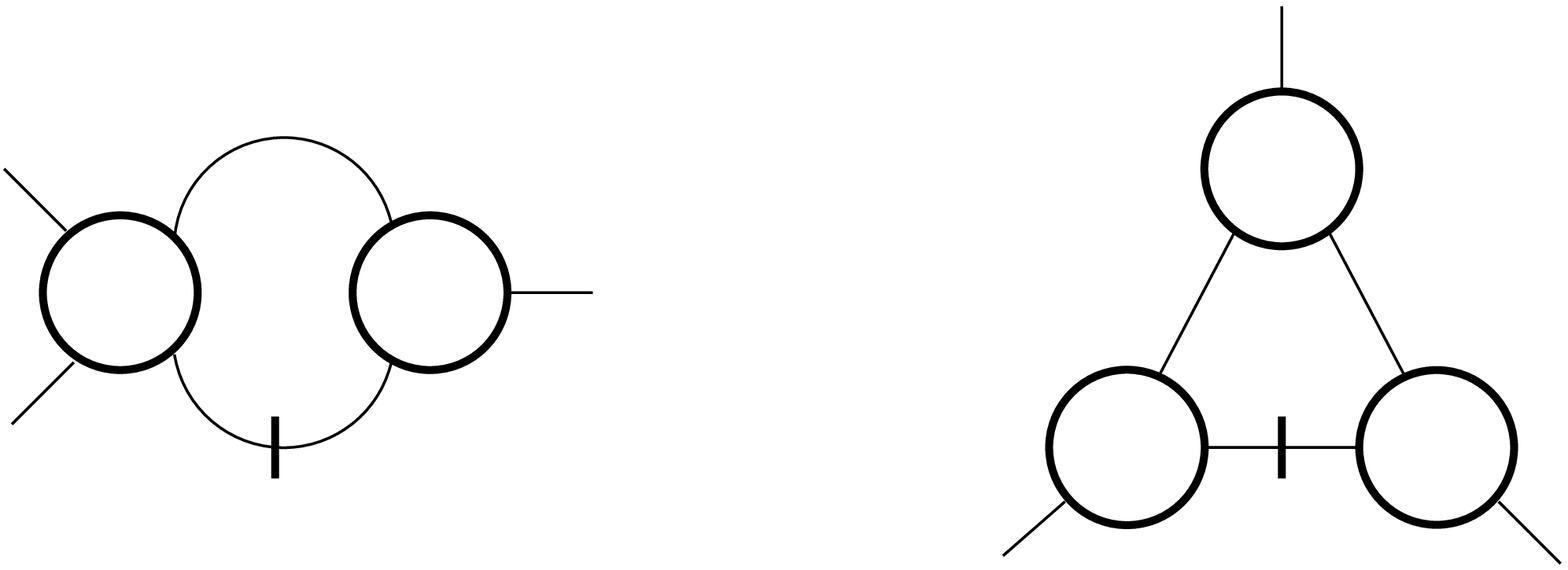,height=3.cm,width=10.cm,angle=0}}
\centerline{(d)}
\centerline{\psfig{file=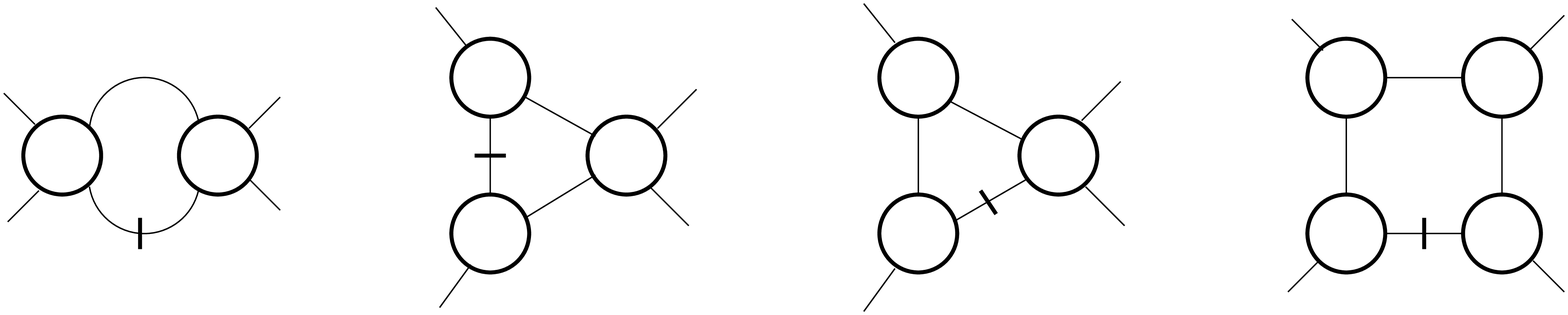,height=3.5cm,width=16.cm,angle=0}}
\centerline{(e)}
\caption{Graphs contributing to the evolution equations of the
coefficient functions of the effective action in the strength of the
disorder. (a): $\dot\Gamma$, (b):$\dot\Gamma_{\hat a}$,
(c): $\dot\Gamma_{\hat a,\hat b}$, (d): $\dot\Gamma_{\hat a,\hat b,\hat c}$,
(e): $\dot\Gamma_{\hat a,\hat b,\hat c,\hat d}$.
The line represents the propagator $F$ and the circle with $n$ legs
the function $\Gamma_{\hat a_1,\ldots,\hat a_n}$. \label{gev}}
\end{figure}

In the rest of this Section we return to the non-interacting case
and consider two different approximations for the evolution equation.
First we use the ansatz
\be\label{rgansatz}
\Gamma[\rho]=\Gamma+\Gamma_{\hat a}\rho_{\hat a}
+\hf\Gamma_{\hat a,\hat b}\rho_{\hat a}\rho_{\hat b}
+{1\over3!}\Gamma_{\hat a,\hat b,\hat c}\rho_{\hat a}\rho_{\hat b}\rho_{\hat c}
+{1\over4!}\Gamma_{\hat a,\hat b,\hat c,\hat d}
\rho_{\hat a}\rho_{\hat b}\rho_{\hat c}\rho_{\hat d}
\ee
where the combined 'super-index' $\hat a=(\alpha,x_a)$ has been introduced
and summation/integration is assumed for double indices.
The inverse of
\be
{\delta^2\Gamma[\rho]\over\delta\rho_{\hat a}\delta\rho_{\hat b}}
=(F^{-1})_{\hat a,\hat b}+\Gamma^{(3)}_{\hat a,\hat b}[\rho]
+\Gamma^{(4)}_{\hat a,\hat b}[\rho]
\ee
where
\bea
(F^{-1})_{\hat a,\hat b}&=&\Gamma_{\hat a,\hat b}\nonu
\Gamma^{(3)}_{\hat a,\hat b}[\rho]
&=&\Gamma_{\hat a,\hat b,\hat c}\rho_{\hat c}\nonu
\Gamma^{(4)}_{\hat a,\hat b}[\rho]
&=&\hf\Gamma_{\hat a,\hat b,\hat c,\hat d}\rho_{\hat c}\rho_{\hat d}
\eea
on the right hand side can be obtained by the Neumann-series,
\be
\dot\Gamma[\rho]=-{g\over2}\int_\mb{x}\left(F
-F\cdot(\Gamma^{(3)}[\rho]+\Gamma^{(4)}[\rho])\cdot F
+F\cdot(\Gamma^{(3)}[\rho]+\Gamma^{(4)}[\rho])\cdot F
\cdot(\Gamma^{(3)}[\rho]+\Gamma^{(4)}[\rho])\cdot F-\cdots\right)
_{(0,0,\mb{x}),(0,0,\mb{x})}.
\ee
The identification of the coefficients of the same $\rho$ powers
gives the equations
\bea\label{reszlgev}
\dot\Gamma&=&-{g\over2}\int_\mb{x}F_{(0,0,\mb{x}),(0,0,\mb{x})}\nonu
\dot\Gamma_{\hat a}\rho_{\hat a}&=&{g\over2}\int_\mb{x}
\left(F\cdot\Gamma^{(3)}[\rho]\cdot F\right)_{(0,0,\mb{x}),(0,0,\mb{x})}\nonu
\dot\Gamma_{\hat a,\hat b}\rho_{\hat a}\rho_{\hat b}
&=&{g\over2}\int_\mb{x}\left[F\cdot\left(\Gamma^{(4)}[\rho]
-\Gamma^{(3)}[\rho]\cdot F\cdot\Gamma^{(3)}[\rho]\right)\cdot F
\right]_{(0,0,\mb{x}),(0,0,\mb{x})}\nonu
\dot\Gamma_{\hat a,\hat b,\hat c}\rho_{\hat a}\rho_{\hat b}\rho_{\hat c}
&=&-{g\over2}\int_\mb{x}\left[F\cdot\left(
2\Gamma^{(4)}[\rho]\cdot F\cdot\Gamma^{(3)}[\rho]
-\Gamma^{(3)}[\rho]\cdot F\cdot\Gamma^{(3)}[\rho]\cdot F
\cdot\Gamma^{(3)}[\rho]\right)\cdot F\right]_{(0,0,\mb{x}),(0,0,\mb{x})}\nonu
\dot\Gamma_{\hat a,\hat b,\hat c,\hat d}
\rho_{\hat a}\rho_{\hat b}\rho_{\hat c}\rho_{\hat d}
&=&-{g\over2}\int_\mb{x}\biggl[F\cdot\biggl(
\Gamma^{(4)}[\rho]\cdot F\cdot\Gamma^{(4)}[\rho]
-\Gamma^{(3)}[\rho]\cdot F\cdot\Gamma^{(4)}[\rho]\cdot F\cdot\Gamma^{(3)}[\rho]\\
&&-2\Gamma^{(4)}[\rho]\cdot F\cdot\Gamma^{(3)}[\rho]\cdot F \cdot\Gamma^{(3)}[\rho]
+\Gamma^{(3)}[\rho]\cdot F\cdot\Gamma^{(3)}[\rho]\cdot F
\cdot\Gamma^{(3)}[\rho]\cdot F\cdot\Gamma^{(3)}[\rho]\biggr)\cdot F
\biggr]_{(0,0,\mb{x}),(0,0,\mb{x})},\nonumber
\eea
shown on Fig. \ref{gev}. It is easy to understand the structure of these
equations. They express the principle that an infinitesimal
change of the strength of the disorder, $g\to g+\delta g$, can be
realized in the order $\ord{\delta g}$ by visiting each
correlation line of the graphs and change its coefficient, represented
by the cross on the correlation lines of the graphs. The equations
express this process in a hierarchical manner, in terms of the
one particle irreducible vertices.

The resummation of the perturbation series, implicit in the solution
of the evolution equation, can easily be seen by iteration,
\be\label{iterat}
W^{(n+1)}_\lambda[\sigma]=W^{(n)}_0[\sigma]
+{g\over2}\int_0^\lambda d\lambda'\int_\mb{p}
{\delta^2W^{(n)}_{\lambda'}[\sigma]\over\delta\sigma_{0,\omega=0,\mb{-p}}
\delta\sigma_{0,\omega=0,\mb{-p}}},
\ee
where each step increases the order of the interactions/correlations 
present by one and
the integration over $\lambda'$ inserts the necessary symmetry factors.
In the numerical integration of the evolution equation each step
$\lambda\to\lambda+\delta\lambda$ increases the order of the 
interactions/correlations
by one and all graphs are resummed as $\delta\lambda\to0$. Naturally
this resummation is performed in the given truncation of the
irreducible vertex functions in the momentum space. In particular,
the the self-energy
insertions, the ladder and crossed diagrams of a particle-hole loop
\cite{langpi,ww,welo} are resummed among other contributions by the first term
on the right hand side of the third equation in \eq{reszlgev}.

\subsection{Gradient expansion}
We consider now the ansatz
\be\label{geffact}
\Gamma[\rho]=\int_x\left[\rho_{\alpha,x}
V_{\alpha,\beta}(\rho_x,-i\partial_x)\rho_{\beta,x}-U(\rho_x)\right]
\ee
for the non-interactive case where the space-time derivatives act
on the function $\rho_{\beta,x}$ only,
\be
V_{\alpha,\beta}(\rho,p)=\sum_AV_{\alpha,A,\beta}(\rho)p^A.
\ee
The combined index $A=(\nu_0,\nu_1,\nu_2,\nu_3)$ defines the partial derivative
$\partial^A=\partial_0^{\nu_0}\partial_1^{\nu_1}\partial_2^{\nu_2}\partial_3^{\nu_3}$.
Since static disorder appears through the $\rho_0$ dependence
in the evolution equation we retain the $\ord{(\rho_j)^2}$ terms
only which are needed for the conductivity. Therefore we shall use
the expression
\bea\label{noncov}
\Gamma[\rho]&=&\int_z\biggl[\hf\rho_{0,z}
\Gamma^{tt}(\rho_{0,z},\sign(Imi\partial_0)i\partial_0,-\Delta)\rho_{0,z}\nonu
&&+\hf\rho_{0,z}\Gamma^{ts}(\rho_{0,z},\sign(Imi\partial_0)i\partial_0,-\Delta)\partial_k\rho_{k,z}
+\hf\rho_{k,z}\Gamma^{ts}(\rho_{0,z},\sign(Imi\partial_0)i\partial_0,-\Delta)\partial_k\rho_{0,z}\nonu
&&+\hf\rho_{j,z}\left(\Gamma^T(\rho_{0,z},\sign(Imi\partial_0)i\partial_0,-\Delta)T_{j,k}(\partial)
+\Gamma^L(\rho_{0,z},\sign(Imi\partial_0)i\partial_0,-\Delta)L_{j,k}(\partial)\right)\rho_{k,z}
-U(\rho_z)\biggr]
\eea
with
\be
U(\rho)=-\hf\Gamma^t\rho_j^2+V(\rho_0)
\ee
and
\be
L(\partial)={\partial\otimes\partial\over\Delta},~~~T(\partial)=1-L(\partial).
\ee
The coefficient functions satisfy the relations
$\Gamma^{ts}(\partial)=-\Gamma^{ts}(-\partial)$,
$\Gamma^{tt}(\partial)=\Gamma^{tt}(-\partial)$, $\Gamma^T(\partial)=\Gamma^T(-\partial)$,
$\Gamma^L(\partial)=\Gamma^L(-\partial)$. The effective action
contains derivatives in arbitrary high order the truncation is
that they act on a single $\rho$ only.
The computation of the second functional derivative used in the
evolution equation is sketched in Appendix \ref{gradexpa}.

The evolution of the functions parameterizing the effective action
can the easiest be obtained by writing
$\rho_{\alpha,x}=\rho_\alpha+\delta\rho_{\alpha,x}$ with $\int_x\delta\rho_{\alpha,x}=0$
and expanding $\Gamma[\rho+\delta\rho]$ in $\delta\rho$ up to terms
$\ord{\delta\rho^2}$. We shall use the notation
$\nabla^j_k=\rho_0\partial^{j+1}_{\rho_0}+k\partial^{j}_{\rho_0}$ with the property
$\partial_{\rho_0}\nabla^j_k=\nabla^{j+1}_{k+1}$, the $\gamma$-functions
$\gamma^{tt}_p=\Gamma^t+\nabla^0_1\Gamma^{tt}(p)$,
$\gamma^{ts}_p=\nabla^0_2\Gamma^{ts}(p)$,
$\gamma^L_p=\Gamma^s+\Gamma^L(p)$,
$\gamma^T_p=\Gamma^s+\Gamma^T(p)$,
and write the left hand side of the evolution equation as
\be
\dot\Gamma[\rho]+\int_p\left\{\hf\delta\rho_{0,-p}\dot\gamma^{tt}\delta\rho_{0,p}
+i\delta\rho_{0,-p}\dot\gamma^{ts}(p)p_j\delta\rho_{j,p}
+\hf\delta\rho_{j,-p}\left[\dot\gamma^T(p)T_p+\dot\gamma^L(p)L_p\right]
\delta\rho_{k,p}\right\}
\ee
The right hand side gives
\bea
&&-{g\over2}\int_p\delta_{p_0,0}V(\Gamma^{(0)-1})_{-p,p}^{0,0}
+{g\over4}\int_p\delta_{p_0,0}(\Gamma^{(0)-1})_{-p,p}^{0,\beta}
\int_r\delta\rho_{\gamma,-r}\delta\rho_{\rho,r}\Gamma^{\beta,\kappa,\gamma,\rho}_{-p,p,-r,r}
(\Gamma^{(0)-1})^{\kappa,0}_{-p,p}\nonu
&&-{g\over2}\int_p\delta_{p_0,0}(\Gamma^{(0)-1})_{-p,p}^{0,\beta}
\int_r\delta\rho_{\gamma,-r}\delta\rho_{\sigma,r}\Gamma_{-p,p+r,-r}^{\beta,\rho,\gamma}
(\Gamma^{(0)-1})^{\rho,\kappa}_{-p-r,p+r}\Gamma_{-p-r,p,r}^{\kappa,\mu,\sigma}
(\Gamma^{(0)-1})^{\mu,0}_{-p,p}+\ord{\delta\rho^3},
\eea
where $F^{\rho,\kappa}_r$, $\Gamma^{\beta,\kappa,\gamma,\rho}_{-t,t,-r,r}$ and
$\Gamma_{-r,t,r-t}^{\kappa,\mu,\sigma}$ are recorded in Appendix
\ref{gradexpa}.
The second term resums the impurity correlations
attached to the particle-hole loop, the self-energy insertions
together with the vertex corrections. The third term contributes
at vanishing energy only and will be neglected.

The $\ord{\delta\rho^0}$ part of the evolution equation is
\be\label{uevol}
\dot U={g\over2}\int_\mb{p}\Gamma^{-1(0,0)})_\mb{p}=
{g\over2}\int_\mb{p}{\gamma^L_p\over\gamma^{tt}_p\gamma^L_p
+\hat\mb{p}^2(\gamma^{ts}_p)^2}
\ee
The $\ord{\delta\rho^2}$ part gives
\bea\label{gevol}
\dot\gamma^{tt}_r&=&{g\over2}\int_\mb{p}{[\partial^2_{\rho_0}\gamma^{tt}_p
+\partial^2_{\rho_0}\gamma^{tt}_r](\gamma^L_p)^2
\over[\gamma^{tt}_p\gamma^L_p
+\hat\mb{p}^2(\gamma^{ts}_p)^2]^2}\nonu
\dot\gamma^{ts}_r
&=&{g\over4}\partial_{\rho_0}^2\gamma^{ts}_r\int_\mb{p}
{(\gamma^L_p)^2\over[\gamma^{tt}_p\gamma^L_p
+\hat\mb{p}^2(\gamma^{ts}_p)^2]^2}\nonu
\dot\gamma^T_r&=&{g\over4}\partial^2_0\gamma^T_r
\int_\mb{p}{(\gamma^L_p)^2\over[\gamma^{tt}_p\gamma^L_p
+\hat\mb{p}^2(\gamma^{ts}_p)^2]^2}\nonu
\dot\gamma^L_r&=&{g\over4}\partial^2_0\gamma^L_r
\int_\mb{p}{(\gamma^L_p)^2\over[\gamma^{tt}_p\gamma^L_p
+\hat\mb{p}^2(\gamma^{ts}_p)^2]^2}.
\eea
One can verify as a consistency check that the $\ord{r^0}$ parts of
the equations for $\gamma^L$ and $\gamma^T$ agree with \eq{uevol}.
Eqs. \eq{gevol} have the form of a one-dimensional diffusion
equation with $(\rho_0,g)$ as 'space-time' coordinates with a
'space'-dependent diffusion constant given by the one-loop
integrals.

The evolution equations need two boundary conditions in $\rho_0$.
One can be obtained by noting that the Green functions 
$\tD^{\mu,\nu}$ are symmetrical under charge conjugation
$\mu\to6\hbar^2/ma^2-\mu$. Therefore $\partial_{\rho_0}\gamma=0$
will be chosen as a boundary condition at half-filling,
$\mu=3\hbar^2/ma^2$, $\rho_0=1/2a^3$ and the density will be
restricted into the interval $0\le\rho_0\le1/2a^3$. The other boundary 
condition is imposed at the lowest density point which is chosen at
sufficiently small chemical potential in such a manner that only one particle 
is found in the system. By assuming that the Green functions increase
linearly with the density in this low density regime the boundary condition
$\partial_{\rho_0}^2\gamma=2\gamma/\rho_0^2$ will be imposed.

\section{Initial conditions}\label{initcos}
After having obtained the evolution equations \eq{uevol}
we need their initial conditions, imposed in the perturbative regime.
One might choose either $\lambda=0$ and use the free effective action
or take a small but non-vanishing $\lambda$ and compute the initial
conditions perturbatively. For a computation carried out at vanishing
frequency and on infinitely large system one has to start with small,
non-vanishing strength of disorder. The reason is that we integrate
a differential equation in this scheme and have to assume that the
$\lambda$-dependence is differentiable. But non-interacting particles
move ballistically in the absence of disorder and an arbitrarily weak
disorder already induces important changes in the transport coefficients.
In fact, the system becomes localised in one or two dimensions at arbitrarily
weak disorder. Sufficiently dilute three dimensional systems should be 
localised at weak impurity field and the construction of the initial 
conditions at weak disorder is rather non-trivial \cite{elk}.

In a more formal language
one may say that the difference between the initial conditions imposed
at $\lambda=0$ and $\lambda\approx0$ lies in the way time reversal symmetry is
broken. In order to arrive at finite
conductivity one needs the breakdown of the time reversal
and the space-translation invariance. Neither of these breakdowns
takes place in the absence of interactions between the electrons
and their environment. The interaction with static impurities
breaks translation invariance. The
interactions break the time reversal invariance dynamically
in infinite volume. The advantage of the perturbative initial
conditions imposed at small but non-vanishing $\lambda$
is that they contain the desired symmetry breaking pattern.

But we have to bear in mind that the numerical solution of
the evolution equation can be obtained for a finite system only.
Furthermore, in order to avoid the singular chemical potential
dependence in finite systems it is necessary
to perform the computation at finite frequency and to study
the limit $\omega\to0$ numerically. The $\lambda$-dependence of
the Green functions of a finite system or at finite frequency is
regular at $\lambda=0$ and no difficulty is expected during
the integration of the evolution equation.
Therefore the initial conditions will be constructed in the absence of
impurities, at $\lambda=0$. The time reversal invariance is then
broken externally, by the non-vanishing frequency and the transition
from the ballistic to the diffusive regime will be induced smoothly
during the evolution.

It is worthwhile noting that the issue of localization of
non-interacting particles is far from being a one-particle problem
and the density-current functional is a highly non-trivial
object even for non-interacting electrons without
disorder, at $\lambda=0$. This is because the Pauli blocking
of the occupied states represents a strong topological
many-particle interaction which obeys no small parameter.
This circumstance is crucial in expecting a singularity
in the ground state in function of the disorder strength.

\subsection{Current and density Green functions}
The first step is the construction of the generator
functional for the connected Green functions of
the density and the current. For this end
we write $\sigma_{\alpha,x}=\sigma_\alpha+\delta\sigma_{\alpha,x}$
where $\int_x\delta\sigma_{\alpha,x}=0$ and find the
following generator functional
\be
W[\sigma+\delta\sigma]=-i\hbar\tr\log\left[G^{-1}+\sigma\br+\delta\sigma\br\right]
=\hbar\tr\log D^{-1}+\rho^*\cdot\delta\sigma
-\hf\delta\sigma\cdot\tD\cdot\delta\sigma+\ord{\delta\sigma^3}
\ee
where $D^{-1}_q=G^{-1}_q+\sigma\br$,
$\rho^*_{\alpha,z}=-i\hbar\tr\left[D\cdot\oper_{\alpha,z}\right]=\rho^*_\alpha$, and
$\tD_{(\alpha,x),(\beta,y)}=-i\hbar
\tr\left[D\cdot\oper_{\alpha,x}\cdot D\cdot\oper_{\beta,x}\right]$.
Due to rotational invariance one has the form
\be
\tD^{\mu,\nu}_{\omega,\mb{q}}=\pmatrix{\tD^{tt}(\omega,\mb{q}^2)&
i\hat\mb{q}e^{-{i\over2}a\mb{q}}\tD^{ts}(\omega,\mb{q}^2)\cr
i\hat\mb{q}e^{{i\over2}a\mb{q}}\tD^{ts}(\omega,\mb{q}^2)&
\tD^{j,k}(\omega,\mb{q})}
\ee
with $\tD^{j,k}(\omega,\mb{q})=T^{j,k}(\mb{q})\tD^T(\omega,\mb{q}^2)+L^{j,k}(\mb{q})\tD^L(\omega,\mb{q}^2)$,
$\hat\mb{q}={2\over a}\sin{a\mb{q}\over2}$,
$L=\hat\mb{q}\otimes\hat\mb{q}/\hat\mb{q}^2$
and $T=1-L$. But on a lattice with finite lattice spacing there
is no rotational symmetry and the transverse piece becomes more
involved. This complication will be neglected below. A more detailed
expression for the propagator $\tD^{\mu,\nu}$ is presented in
Appendix \ref{pertexp}.

\subsection{Variable transformation}
The second step in constructing the density-current functional
for the initial condition is the change of variables $\sigma\to\rho$.
For this end we have to invert the relation
\be
\rho_{\alpha,x}=\left(\rho_\alpha+\delta\rho_{\alpha,x}\right)
=-i\hbar\tr\left[{1\over D^{-1}+\sigma\br+\delta\sigma\br}\cdot\oper_{\alpha,x}\right]
=i\hbar\tr\left[D\cdot\oper_{\alpha,x}\right]
+i\hbar\tr\left[D\cdot\delta\sigma\br\cdot D
\cdot\oper_{\alpha,x}\right]+\ord{\rho^2}.
\ee
We write $\mu+\rho_0\to\mu$ set $\rho_j=\sigma_j=0$ in the absence
of Coulomb interaction and find
\be
\delta\sigma=-\tDi\cdot\delta\rho
\ee

\subsection{Effective action}
Finally, the effective action can be written as
\be\label{ansgrexp}
\Gamma[\rho]=i\hbar\tr\log D^{-1}+\sigma\cdot\rho
-\hf\delta\rho\cdot\tDi\cdot\delta\rho+\ord{\delta\rho^3}.
\ee
The identification of the part $\ord{\delta\rho^0}$ yields
\be\label{homgr}
U=-\sigma_\alpha\rho_\alpha-{i\hbar\over V}\tr\log D^{-1}.
\ee
The $\ord{\delta\rho^2}$ order, the second functional derivative gives
\be
\pmatrix{\gamma^{tt}_q&i\hat\mb{q}e^{-{i\over2}a\mb{q}}\gamma^{ts}_q\cr
i\hat\mb{q}e^{{i\over2}a\mb{q}}\gamma^{ts}_q&T\gamma_q^T+L\gamma_q^L}
=-\pmatrix{\tD^{tt}&i\hat\mb{q}e^{-{i\over2}a\mb{q}}\tD^{ts}\cr
i\hat\mb{q}e^{{i\over2}a\mb{q}}\tD^{ts}&T\tD_T+L\tD_L}^{-1}
=-\pmatrix{{\tD_L\over D}&-i\hat\mb{q}e^{-{i\over2}a\mb{q}}{\tD^{ts}\over D}\cr
-i\hat\mb{q}e^{{i\over2}a\mb{q}}{\tD^{ts}\over D}&{1\over\tD_T}T+{\tD^{tt}\over D}L},
\ee
where
\be
D(\omega,\mb{q}^2)=\tD^{tt}(\omega,\mb{q}^2)\tD^L(\omega,\mb{q}^2)
+\hat\mb{q}^2(\tD^{ts}(\omega,\mb{q}^2))^2.
\ee
Therefore
\bea
\gamma^{tt}(\omega,\mb{q}^2)&=&-{\tD^L(\omega,\mb{q}^2)\over D(\omega,\mb{q}^2)}\nonu
\gamma^T(\omega,\mb{q}^2)&=&-{1\over D^T(\omega,\mb{q}^2)}\nonu
\gamma^L(\omega,\mb{q}^2)&=&-{\tD^{tt}(\omega,\mb{q}^2)\over D(\omega,\mb{q}^2)}\nonu
\gamma^{ts}(\mb{q}^2)&=&{\tD^{ts}(\omega,\mb{q}^2)\over D(\omega,\mb{q}^2)}.
\eea

The Euclidean Ward identities require
\be
\tD^{ts}(\omega,\mb{q})=-{1\over\omega}\left(
\tD^{j,k}(\omega,\mb{q})+{1\over m}\delta^{j,k}\tilde\rho^*\right)
\ee
and
\be
\tD^{tt}(\omega,\mb{q}^2)={\hat\mb{q}^2\over\omega}\tD^{ts}(\omega,\mb{q}^2),
\ee
c.f. Appendix \ref{wards}. It is easy to see that these equations
restrict $\gamma^L_p$ to be momentum independent.

\section{Numerical results}\label{numers}
There are two
points where the numerical integration can be made faster.
First, the one-loop integral in the evolution equation requires the knowledge
of the $\gamma$ functions in the whole Brillouin zone. It was assumed
that the $\gamma$
functions depend on $\hat\mb{p}^2$, momentum square only. This assumption
was found to be acceptable in the initial condition. Note in this respect
that $\tD^L$ and $\gamma^L$ remain well defined in a non-isotrop
Brillouine zone. The other improvement on the speed of the algorithm
is the runtime adjustment of the step size $\Delta g$ during the
numerical integration. This was carried out by keeping the
largest relative increment in the differential equations under $1\%$.
Lattices up to size $320^3$ were used but the results presented 
below correspond to the volume $80^3$ unless it is stated otherwise.
The units $a=m=\hbar=1$ were used to express dimensional quantities.
The integration of the evolution equation started at $g=0$ with 
$\Delta g=10^{-6}$ and ended at $g\approx10^5$, where 
$\Delta g\approx10^4$ was used. 60 division
points were used to monitor the density dependence of the
$\gamma$ functions in the interval $2\times10^{-6}<\rho_0<0.5$.
The Euclidean frequency was restricted by the bounds $10^{-5}<\omega<335$.

The Ward identities allow us to express the longitudinal
conductivity as
\be
{\sigma(\omega,\mb{p}^2)\over ie^2}=-\tD^{ts}(\omega,\mb{p}^2)
={\omega\over\mb{p}^2\gamma^{tt}(\omega,\mb{p}^2)+\omega^2\gamma^L(\omega)},
\ee
indicating that the d.c. conductivity in the homogeneous
$\mb{p}=0$ limit corresponds to an I.R. cut-off, a 'mass gap'
in the propagator $\tD^{ts}(\omega,\mb{p}^2)$ and it can only be
non-vanishing due to an $\ord{1/\mb{p}^2}$ singularity in $\gamma^{tt}$
functions ($\gamma^L$ remains finite as $\omega\to0$). A rough,
qualitative picture of what happens as the disorder is increased
is given by the analogy with the diffusion equation. The $\gamma$-functions
of the initial condition have a singular peak at $\rho_0\approx0$.
Therefore the diffusive evolution smears this peak out and
tries to 'fill up' the $\gamma$ functions at large densities
where they take smaller values.
Since the propagator $\tD^{tt}$ is $\ord{1/V}$ for a system of
finite volume $V$ the diffusion is rather slow at low densities.
Therefore the main $g$-dependence will be an increase of the
$\gamma$ functions for intermediate and large densities. As this
increase diffuses 'up' in the density it increases the 1PI functions
and suppresses the conductivity.

\begin{figure}
\centerline{\psfig{file=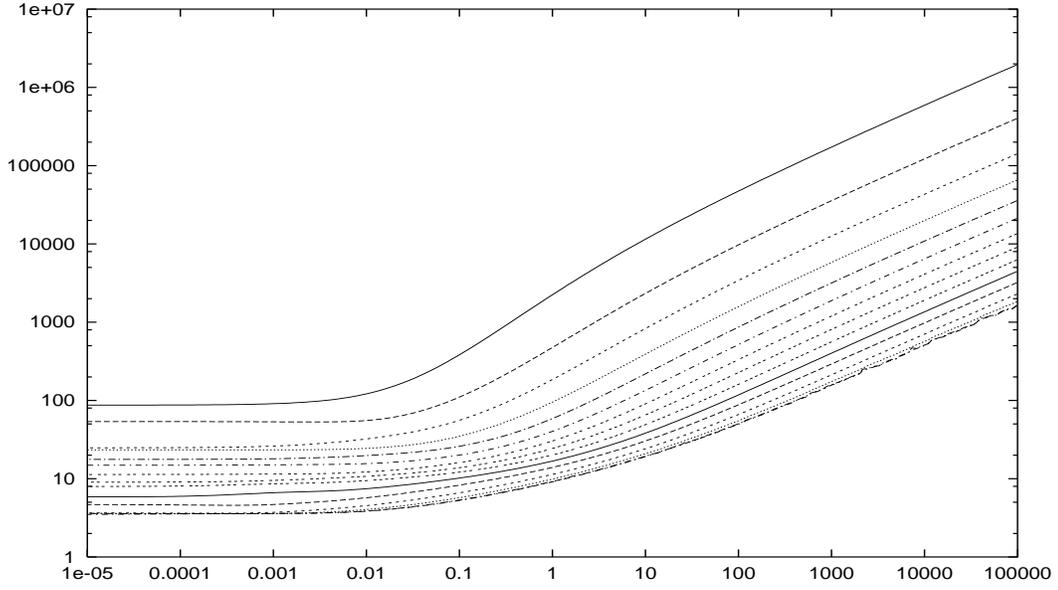,height=8cm,width=14cm,angle=270}}
\centerline{(a)}
\centerline{\psfig{file=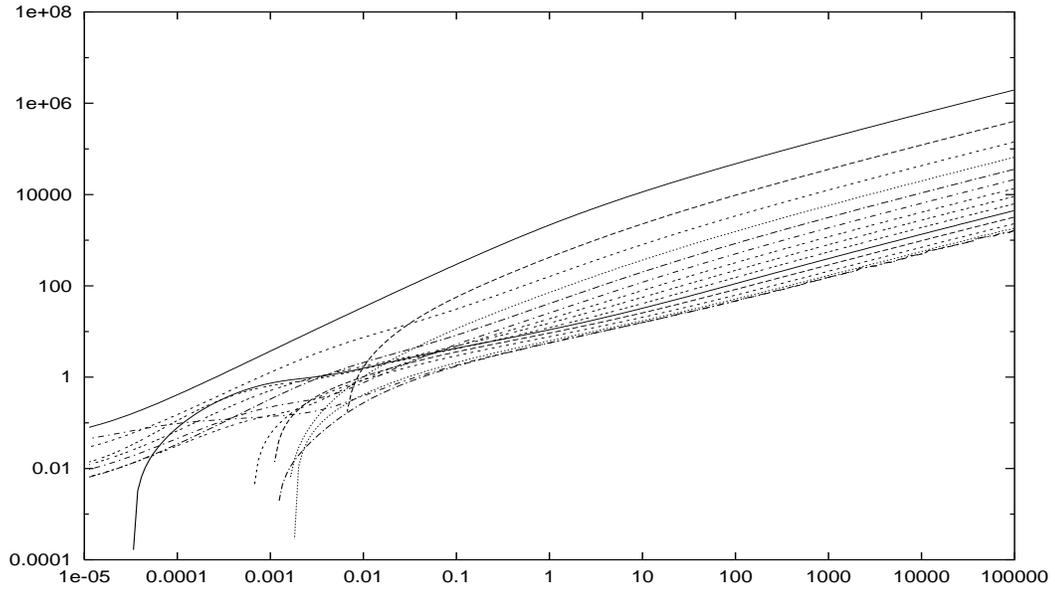,height=8cm,width=14cm,angle=270}}
\centerline{(b)}
\caption{a: Inverse conductivity $\sigma^{-1}(g)$ as the function of
the coupling constant $g$ at $\omega=2\times10^{-5}$.
Different lines correspond to different densities in the
interval $7.6\times10^{-3}\le\rho_0\le0.5$. b: The same as (a) except for
$\Delta\sigma^{-1}=\sigma^{-1}(g)-\sigma^{-1}(0)$.
Both $\sigma^{-1}$ and $\Delta\sigma^{-1}$ are monotonically
decreasing function of the density for $g>1$. The negative
$\Delta\sigma^{-1}$ values were ignored in the logarithmic plot.
\label{cev}}
\end{figure}

Let us start with the $g$-dependence of $\sigma^{-1}(\omega,\mb{p}^2)$,
shown in Fig. \ref{cev} (a) at $\omega=2\times10^{-5}$ and
$\mb{p}=\mb{p}_1=(0,0,\pi/80)$ for densities
$7.6\times10^{-3}<\rho_0<0.5$. The
inverse conductivity displays a plateau at weak disorder
and starts to increase at $g\approx0.1$ only.
In order to see clearer what happens at this crossover point
$\Delta\sigma^{-1}(g)=\sigma^{-1}(g)-\sigma^{-1}(0)$,
the contribution due to the disorder, is shown in Fig. \ref{cev} (b).
For some densities one finds a slight drop in $\sigma^{-1}$ at the
beginning of the evolution, the figure shows the values with
$\Delta\sigma^{-1}(g)>0$ only. We believe that this weak
non-monotonic $g$-dependence is a finite size effect. It is induced
by the evolution equation when the density dependence in the initial 
conditions is distorted by a finite size effect. The latter is
due to the constraint $\Theta(-E_{\mb{q}})$
in the integrand of the zero temperature Green functions which
induces an artificial step function-like chemical potential 
dependence in the initial conditions, \eq{initialc}.
When this finite size effect is ignored the lesson of Fig. \ref{cev}
(b) is that the change of the nature of $\sigma^{-1}$ at $g\approx1$
in Fig. \ref{cev} (a) is not a singularity. What happens is that
the increase of $\sigma^{-1}$ due to disorder becomes
comparable with the initial value and disorder
starts to play a more important role. $\Delta\sigma^{-1}(g)$
plotted for different values of the momentum $\mb{p}$ shows similar
behavior. One finds the scaling $\Delta\sigma^{-1}(g)\approx g^k$
with $k=1$ for $g<<1$ and $k=1/2$ as $g>>1$ without
any non-analicity in the $g$ dependence. The behavior 
$\Delta\sigma^{-1}(g)\approx g$ characterizes weak disorder.
It is easy to see that the condition of the simplification that 
the $g$-dependence 
generated by the evolution equations \eq{gevol} factorizes as an overall 
multiplicative factor is just the asymptotic scaling
$\Delta\sigma^{-1}(g)\approx\sqrt{g}$. The possibility of
any sudden transition at $\rm{p}=0$ developing in the
thermodynamical and the $\omega\to0$ limit between the weak and strong 
disorder region is still hidden in the momentum dependence, the issue
we turn to now.

\begin{figure}
\centerline{\psfig{file=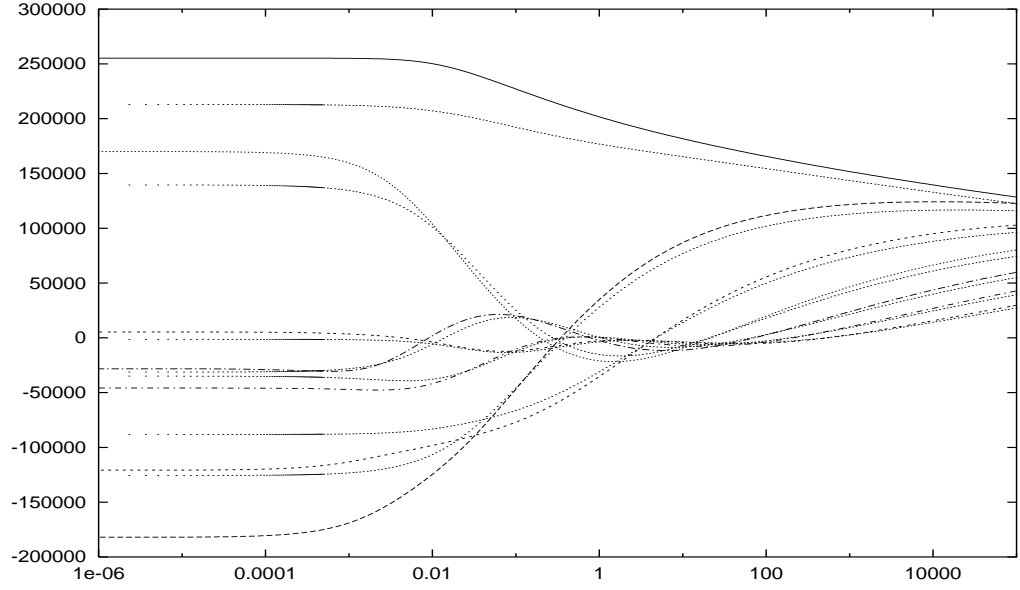,height=8cm,width=14cm,angle=270}}
\centerline{(a)}
\centerline{\psfig{file=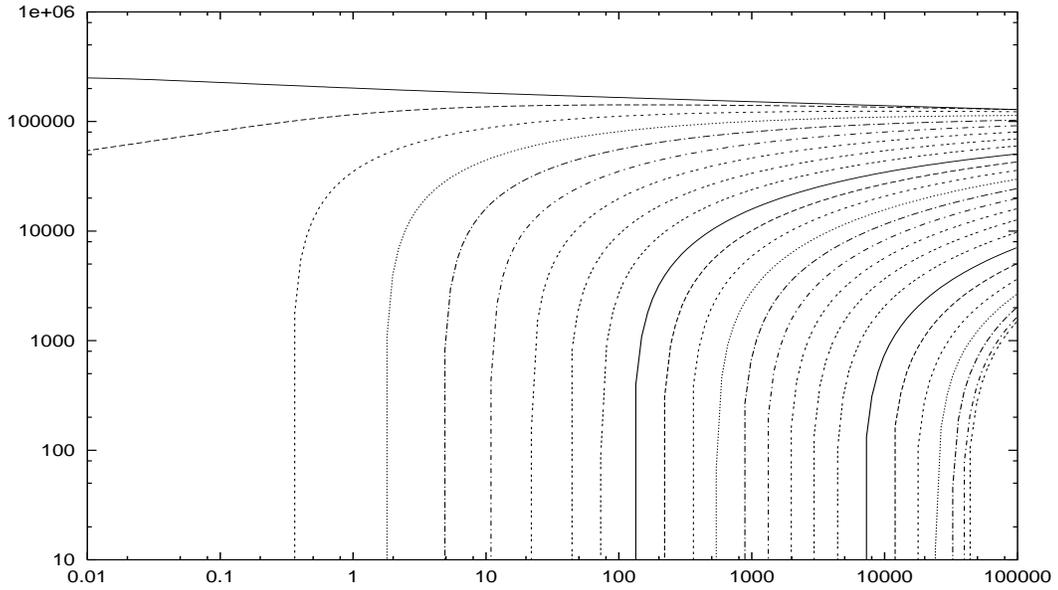,height=8cm,width=14cm,angle=270}}
\centerline{(b)}
\caption{(a): Inverse conductivity extrapolated to
vanishing momentum at $\omega=0.0002$ as the function of $g$
for $0.0171<\rho_0<0.19$. The solid and dotted lines correspond
to lattices $80^3$ and $160^3$, respectively. The finite size
dependence is stronger for $\sigma^{-1}<0$. (b):
The extrapolated $\sigma^{-1}$ as the function of $g$ for
densities $0.0171<\rho_0<0.5$. Only the values $\sigma^{-1}>0$
are shown and $\sigma^{-1}$ is a monotonically
decreasing function of $\rho_0$ in this regime.\label{extr}}
\end{figure}

\begin{figure}
\centerline{\psfig{file=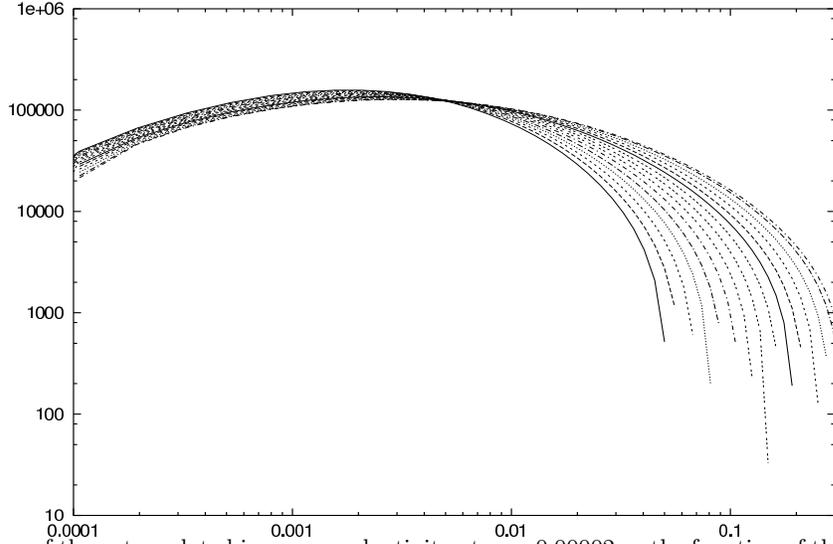,height=8cm,width=12cm,angle=270}}
\caption{The positive values of the extrapolated
inverse conductivity at $\omega=0.00002$ as the function of the
density for different strength of disorder,
$180<g<4\times10^4$. $\sigma^{-1}$ is increasing monotonically
with $g$ at high densities. \label{extrd}}
\end{figure}

In order to explore the I.R. regime the extrapolation of the
conductivity to $\mb{p}=0$ was performed by fitting the $\mb{p}^2$
dependence of $\sigma^{-1}(\omega,\mb{p}^2)$ by a polynomial
up to $\mb{p}^{12}$. The result is dominated by finite size
effects at weak disorder. In fact, $g=0$, a perfectly ordered
system should have vanishing $\sigma^{-1}$ at $\mb{p}=0$, i.e.
long range Green function $\tD^{ts}$. In more physical terms,
the density of state is a sum of Dirac-delta peaks for finite
systems and one needs rather large volume in order to approach
a continuous density of state which is needed for the Bragg
reflections on the lattice and for finite conductivity. The
dependence of the inverse conductivity on the strength of disorder,
depicted in Fig. \ref{extr} (a) supports this expectation.
Furthermore one can see that at low densities where the
density of states is lower the conductivity is strongly
suppressed and $\sigma^{-1}$ is large and positive for any $g$.
When the negative parts of $\sigma^{-1}$ are discarded as in
Fig. \ref{extr} (b) one sees the sudden drop of the
conductivity as the strength of disorder is increased.
$\sigma^{-1}$ is plotted against the density in Fig.
\ref{extrd} showing that the density dependence of
the a.c. conductivity is strongly suppressed for strong disorder.

\begin{figure}
\centerline{\psfig{file=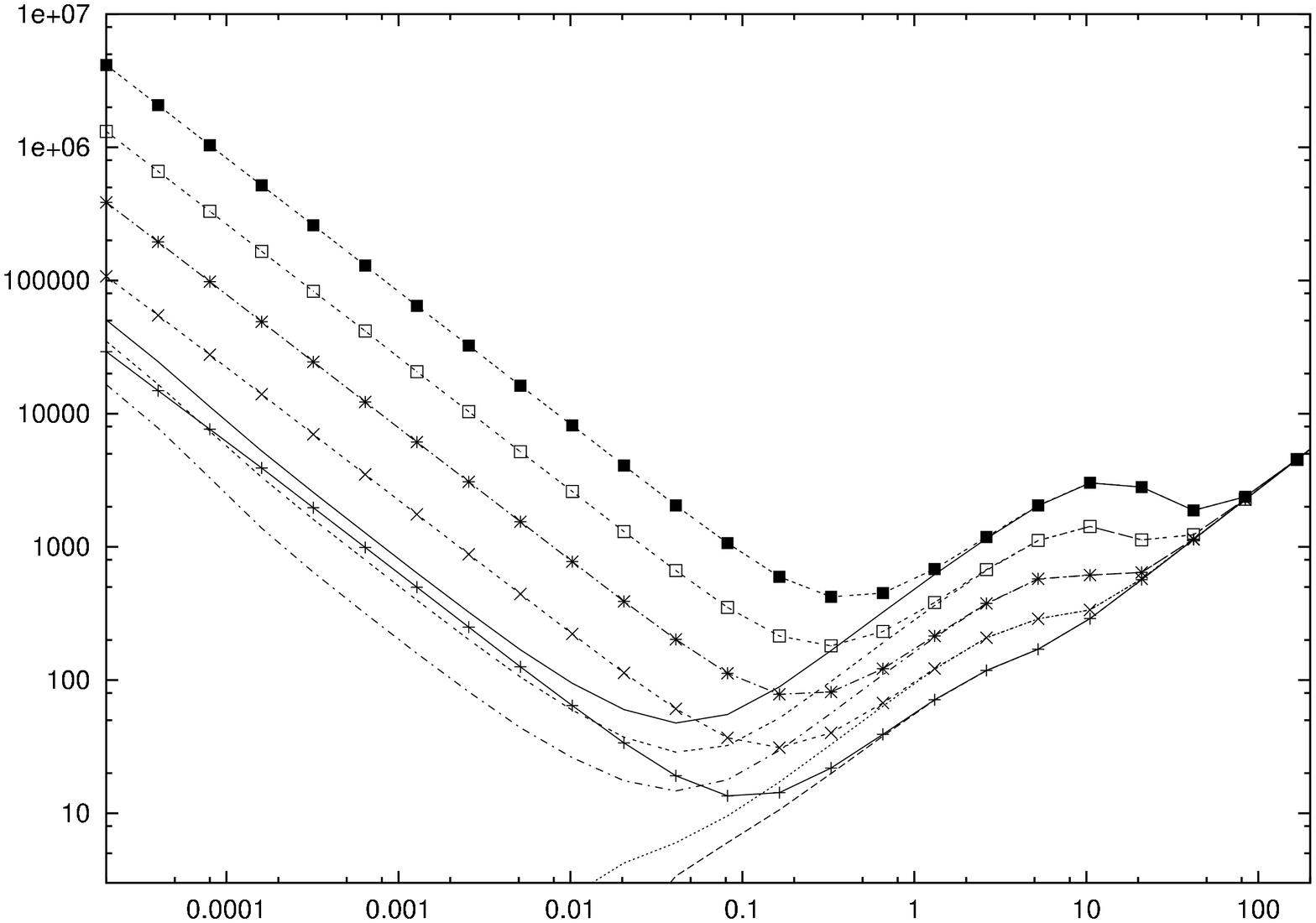,height=6.cm,width=12cm,angle=270}}
\centerline{(a)}
\centerline{\psfig{file=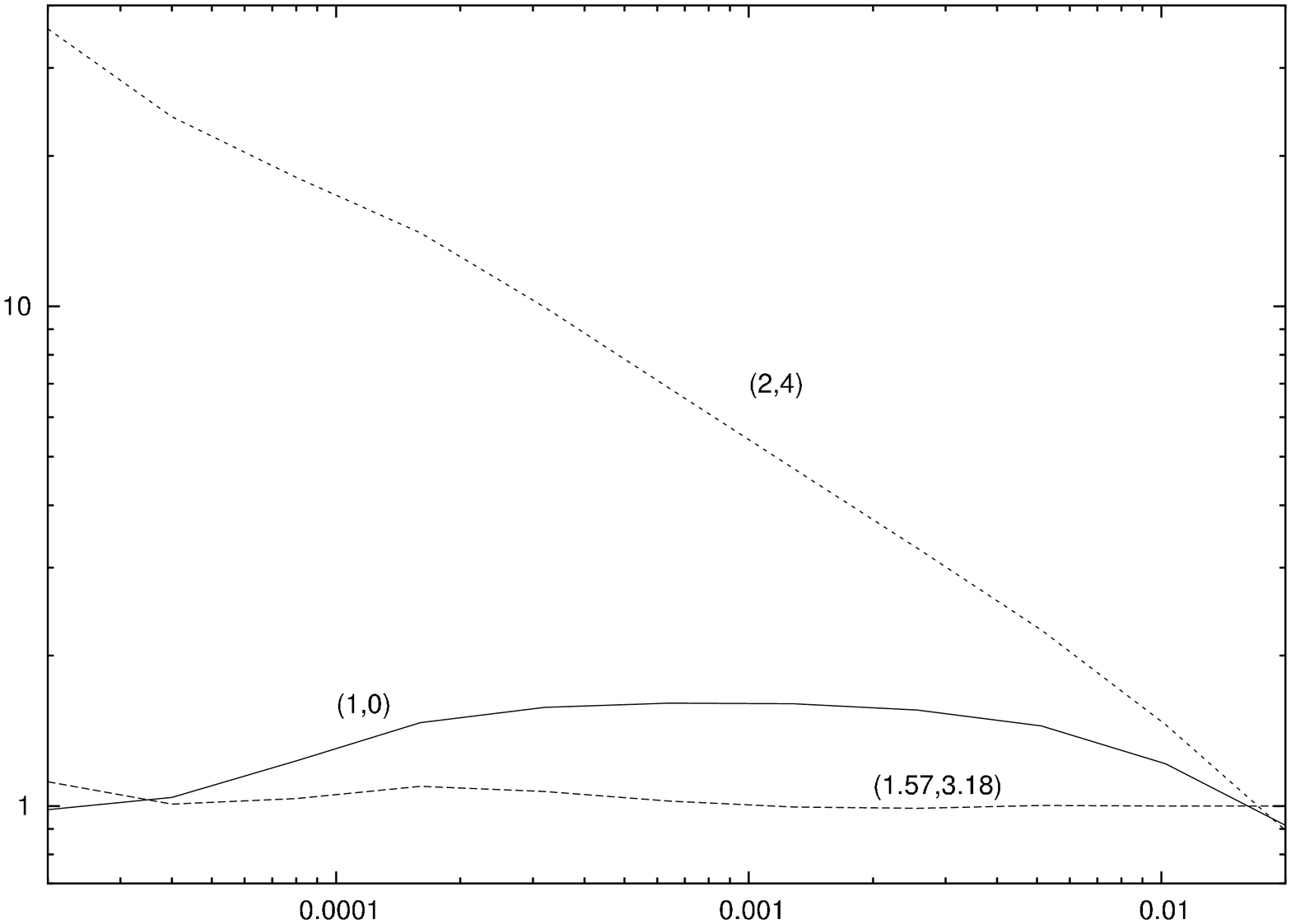,height=6.cm,width=12cm,angle=270}}
\centerline{(b)}
\eject
\caption{(a):The solid lines show the positive part of the
inverse conductivity extrapolated to
vanishing momentum as the function of the frequency for
$\rho_0=0.11$ and $g=10,~10^2,~10^3~,10^4$, and $10^5$.
Lines with symbols correspond to the inverse conductivity with
momentum $\mb{p}=\mb{p}_1=(0,0,\pi/80)$. $\sigma^{-1}$ is a
monotonically increasing function of $g$.
(b): The ratio $\sigma^{-1}_f/\sigma^{-1}$ plotted as the function of $\omega$
in the localised regime, at $g=10^2$, for different choices
of $(\kappa,\eta)$ shown beside the curves. Notice that $\sigma^{-1}$
changes by three order of magnitude in the frequency regime shown
in this plot. \label{ent}}
\end{figure}

For the identification of the sudden increase of $\sigma^{-1}$ with
the onset of the localized phase we have to perform the limit $\omega\to0$.
The frequency dependence of $\sigma^{-1}$ is shown in Fig. \ref{ent} (a).
One recovers the usual $\sigma\approx1/\omega$ behavior at any
momentum $\mb{p}$ at high frequencies except a short flattening
around $\omega\approx10$ where the disorder decouples. In fact,
the evolution is suppressed at high frequencies due to the smallness of
propagators in the initial conditions, c.f. \eq{initialc}, and the scaling
$\sigma\approx1/\omega$ is recovered at higher frequencies
with a proportionality constant given approximately by the initial conditions.
As we turn towards low frequencies we find two qualitatively different
behaviors for the conductivity extrapolated to $\mb{p}=0$.
At weaker disorder $\sigma^{-1}$ decreases and becomes negative,
more precisely is dominated by finite size effects and should not be
taken seriously when computed at the present size lattices.
On the contrary, for stronger disorder the computation is free
of finite size effects and $\sigma$ decreases as $\omega\to0$. 
Motivated by the Mott form
\be\label{mott}
\sigma_f=C\left({\omega\over\omega_0}\right)^\kappa
\left|\log{\omega\over\omega_0}\right|^\eta
\ee
with $(\kappa,\eta)=(2,4)$, the values of $\kappa$ $\eta$, $\omega_0$
and $C$ were fitted in Eq. \eq{mott} for strong disorder.
Mott's result alone compares rather
poorly with the numerical results as shown in Fig. \ref{ent} (b). The
simplest choice, $(\kappa,\eta)=(1,0)$ represents a better approximation
but the fit with the results $(\kappa,\eta)=(1.57,3.18)$ shows the
presence of logarithmic corrections and gives a further improvement.

\begin{figure}
\centerline{\psfig{file=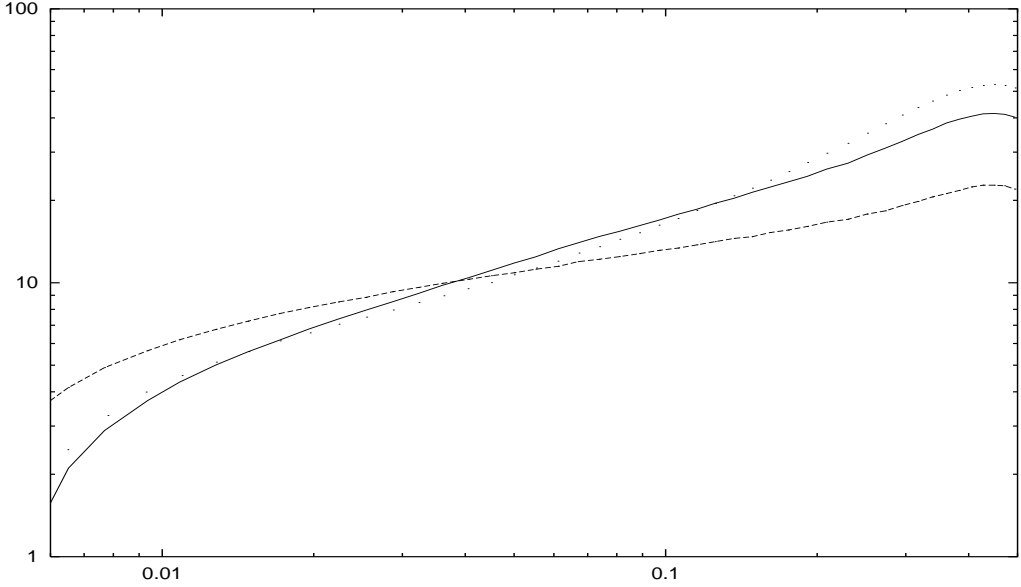,height=8.cm,width=14cm,angle=270}}
\caption{The average strength of the disorder potential
in units of the band width, $\sqrt{g_{cr}}/E_F$ on $80^3$ lattice
as the function of the density for $\omega=0.00512$ (heavy line)
and $0.00002$ (thin line). The dots represent the same
quantity obtained on $320^3$ lattice. The finite size
dependence is invisible on the lower frequency curve.\label{gcr}}
\end{figure}

There is a discontinuity at $g=g_{cr}$ separating finite and
diverging values of $\sigma^{-1}$ at weak and strong disorder, respectively.
The critical strength of the disorder, $g_{cr}$ can be estimated
by finding where $\sigma^{-1}$ changes sign first as $g$ is decreased,
c.f. Fig. \ref{gcr}.
This is how the localization-delocalization transition appears in this scheme.

\section{Conclusions}\label{concl}
A new, non-perturbative scheme is presented in this paper for
the description of disordered system. The basic idea is to express
the effects of an infinitesimal increase of the disorder by changing
the density. No small parameter is required to obtain the evolution
equation but truncation is needed to convert it into a useful numerical
algorithm. We believe that the gradient expansion is a well suited and
consistent truncation scheme for the computation of the transport
coefficients.

There is a rather detailed space-time picture for weak localization
drawn from the partial resummation of the perturbation expansion \cite{welo}.
The mechanism proposed in this paper is more formal, it is a diffusion
process where the density and disorder strength are considered as
space and time variables. The 1PI functions take large values
at low density in the initial conditions. The conductivity decreases
when the disorder is becoming stronger due to the diffusion of the
low density peak of the 1PI functions towards higher densities.
Localization appears as a special, correlated increase of the 1PI functions
at different momentum values.

The numerical results obtained in solving the evolution equation
indicates that weak localization is rather difficult to establish
due to strong finite size effects. In fact, the finite conductivity
arises from an $\ord{1/\mb{p}^2}$ I.R. singularity of a 1PI function
whose reproduction requires unusually large lattices. But strong disorder
suppresses the long range correlations and renders the numerical
solution more reliable.

It was found that the conductivity is vanishing as $\omega\to0$
when the strength of disorder exceeds a certain threshold. But the
fit of the numerical results yields fractional powers of $\omega$ and
$\log\omega$, indicating that several terms contribute in the frequency
interval considered.
It is natural to identify this threshold with the localization transition.

The results presented here are preliminary, obviously more careful
study of the finite size effects is needed to establish contact
with perturbation expansion and to identify the weak localization regime.
Furthermore, a more extended frequency interval must be studied in order to 
identify the low frequency scaling form of the conductivity in the localized 
regime. Finally, the most important step is the inclusion of the Coulomb 
interaction in the numerical solution.

\acknowledgments
I thank J\'anos Hajdu, Shimon Levit and Peter W\"olfle for illuminating
discussions and encouragement.

\eject

\eject

\eject

\appendix
\section{Notations}\label{notations}
We summarize here briefly the notations used in the paper. The
space-time integrals and the Fourier transform are
\be
\int_{-T/2}^{T/2}dt=\int_t,~~
\int d^dx=a^d\sum_x=\int_x,~~\int_xf_xg_x=f\cdot g,~~
\int{d^dp\over(2\pi)^d}={1\over V}\sum_p=\int_p,~~
f_p=\int_xe^{-ipx}f_x,~~~f_x=\int_pe^{ipx}f_p,
\ee
with $x=(\mb{x},t)$, $p=(\mb{p},-\omega)$, and $xp=\mb{x}\mb{p}-t\omega$.
The Dirac-deltas $\delta_{x,y}=a^{-d}a_\tau\delta_{x,y}^K$ and
$\delta_{p,q}=VT\delta_{p,q}^K$ are expressed in terms of the
Kronecker-deltas $\delta_{x,y}^K$ and $\delta_{p,q}^K$.
Identities for the functional derivatives are
\bea
{\delta\over\delta f_x}&=&{\partial\over\partial a^3a_\tau f_x}
=\sum_pe^{-ipx}{\partial\over\partial f_p}
={1\over VT}\sum_pe^{-ipx}{\delta\over\delta f_p}
=\int_pe^{-ipx}{\delta\over\delta f_p},\nonu
{\delta\over\delta f_p}&=&VT{\partial\over\partial f_p}
=\sum_xe^{ipx}{\partial\over\partial f_x}
=\int_xe^{ipx}{\delta\over\delta f_x}.
\eea
Greek and Latin indices will be used for the range $0,1,2,3$ and $1,2,3$,
respectively.

\section{Annealed v.s. quenched evolution}\label{repl}
It is shown that the annealed and the quenched averages
correspond to simple evolution equation. Let us take
a function $f(w)$ and introduce the generator functional
\be
f(W[\sigma])={\int\cd{v}e^{-S_d[v]}f(W_0[\sigma-iv])\over\int\cd{v}e^{-S_d[v]}}
\ee
where
\be
S_d[v]=\hf v\cdot K\cdot v.
\ee
We make the replacement $v\to v-\sigma$ $(M)$, $v\to v-i\sigma$ $(E)$,
i.e. use the equation of motion for $v$ to write
\be
f(W[\sigma])={\int\cd{v}e^{-S_d[v-i\sigma]}f(W_0[-iv])
\over\int\cd{v}e^{-S_d[v]}}.
\ee
We therefore have the relations
\bea
\dot W[\sigma]f'(W[\sigma])&=&-{\int\cd{v}(v-i\sigma)\cdot\dot K\cdot(v-i\sigma)
e^{-\hf(v-i\sigma)\cdot K\cdot(v-i\sigma)}f(W_0[-iv])
\over2\int\cd{v}e^{-\hf v\cdot K\cdot v}}+\hf f(W[\sigma])\tr\dot K\cdot K^{-1},\nonu
{\delta W[\sigma]\over\delta\sigma_x}f'(W[\sigma])
&=&-{\int\cd{v}{\delta S_d[v-i\sigma]\over\delta\sigma_x}
e^{-S_d[v-\sigma]}f(W_0[-iv])\over\int\cd{v}e^{-S_d[v]}},\\
{\delta \over\delta\sigma_y}\left({\delta W^M[\sigma]\over\delta\sigma_x}f'(W^E[\sigma])\right)
&=&-{\int\cd{v}\left((K\cdot(v-i\sigma))_x(K\cdot(v-i\sigma))_y-K_{x,y}\right)
e^{-\hf(v-i\sigma)\cdot K\cdot(v-i\sigma)}f(W^E_0[-iv])
\over\int\cd{v}e^{-\hf v\cdot K\cdot v}}\nonumber
\eea
which allow us to write the evolution equation as
\be
\dot W[\sigma]
=-\hf\tr\dot K^{-1}\cdot\left({\delta^2W[\sigma]\over\delta\sigma\delta\sigma}
+{\delta W[\sigma]\over\delta\sigma}{f''(W[\sigma])\over f'(W[\sigma])}
{\delta W[\sigma]\over\delta\sigma}\right).
\ee

The simplest equation arises from the assumption
\be
f''(W[\sigma])={n\over\hbar}f'(W[\sigma])
\ee
where $n$ is a constant. The solution,
\be
f(x)=\cases{x&$n=0$\cr e^{nx/\hbar}&$n\not=0$}
\ee
corresponds to the quenched average for $n=0$ and to $n$ annealed
replicas when $n\not=0$. The evolution equation reads as
\be\label{replw}
\dot W_q[\sigma]=-\hf\tr\dot K^{-1}\cdot\left(
{\delta^2W_a[\sigma]\over\delta\sigma\delta\sigma}
+{n\over\hbar}{\delta W_a[\sigma]\over\delta\sigma}
{\delta W_a[\sigma]\over\delta\sigma}\right)
\ee
One finds
\be
\dot\Gamma[\sigma]=\hf\tr\dot K^{-1}\cdot\left[
\left({\delta^2\Gamma^E_a[\rho]\over\delta\rho\delta\rho}\right)^{-1}
+{n\over\hbar}\rho\rho\right]
\ee
in terms of the effective action $\Gamma[\sigma]$. The condition
for the applicability of the replica method is the assumption that
the solution of this differential equation not only converges
as $n\to0$ but the $n$-dependence is continuous, i.e.
the limit agrees with the solution obtained with $n=0$.
According to Eq. \eq{replw} $n$ should approaches zero at least with
the inverse volume in the thermodynamical limit.
This requirement, well known from elementary considerations,
may pose problem in phase transitions or when some symmetry is 
broken spontaneously.

\section{Ward identity}\label{wards}
The Kubo formula for the conductivity describes the way external
electromagnetic potential generates the motion of the charges.
One needs special care in using this formalism when the charges
in question are treated in the second quantized formalism. The
point is that one should not use approximations which violate
the particle number. In fact,
an uncontrolled number of particles created by an approximation
renders the transport coefficients divergent when the
IR limit is considered, whatever weak error is committed in the approximation.
It is easy to control the total number of particles by global
symmetries. But the complications start when the particles
carry a conserved charge which can take positive as well negative values.
In this case the conservation
of the algebraic sum of the charge is not enough any more to protect
the transport coefficient against IR divergences. In fact, imagine
an approximation scheme which though preserves the total electric charge
exactly but leads to the creation of an uncontrolled number of neutral
pairs. The conductivity will obviously be IR divergent in such a scheme
due to the polarization cloud generated by the erroneous approximation.
Charged particles are usually handled by gauge theories and the approximations
which leads to the creation of neutral pairs without violating the
charge conservation are those which are invariant under global
symmetry transformations but violate local gauge transformations.

The best way to make sure that a symmetry is preserved is to
check if the equation of motion in the given regularization and approximation
scheme remains symmetrical. In the derivation of the equation of motion,
\eq{eqmo}, then one uses infinitesimal shift of the dynamical field
variables which corresponds to a symmetry transformation.
The action remains invariant and any contribution to the equation
of motion must come from non-symmetrical terms, i.e. non-symmetrical
parts of the dynamics are the only sources contributing to such
a special combination of the equation of motion, called Ward identity.

In our case we generate the shift of the field variables by an
infinitesimal gauge transformation,
\be\label{gtr}
\psi_x\to e^{{i\over\hbar}\epsilon_x}\psi_x.
\ee
Since the electromagnetic fields appear as an external source
only without dynamics they do not participate in this transformation.
Nevertheless they appear in the discussion because the transformation
\eq{gtr} is equivalent with
\be
U_\mu(x)\to U_\mu(x)e^{-{i\over\hbar}a_\mu\nabla_\mu^+\epsilon_x}
\ee
where
\be\label{gtrr}
\nabla^\pm_\mu f_x=\pm{f_{x\pm\hat\mu}-f_x\over a_\mu}
\ee
as long as the dynamics is gauge invariant. The transformation \eq{gtrr}
induces the change
\bea
\delta S&=&-a^3a_\tau\sum_x\nabla^+_\mu\epsilon_x
\psid\cdot\oper_{\mu,x}\cdot\psi+\ord{\epsilon^2},\nonu
\delta\psid\cdot\oper_{0,x}\cdot\psi
&=&-{ia_\tau\over\hbar}\nabla^+_0\epsilon_x
\psid\cdot\oper_{0,x}\cdot\psi+\ord{\epsilon^2},\nonu
\delta\psid\cdot\oper_{j,x}\cdot\psi
&=&{1\over m}\nabla^+_j\epsilon_x\psid\cdot\tilde\oper_{j,x}\cdot\psi
+\ord{\epsilon^2},\nonu
\delta\psid\cdot\tilde\oper_{j,x}\cdot\psi
&=&-{ma^2\over\hbar^2}\nabla^+_j\epsilon_x\psid\cdot\oper_{j,x}\cdot\psi
+\ord{\epsilon^2}.
\eea
Using this transformation as a reparametrization of the functional integral
including the operators $\oper_\mu$ and $\tilde\oper_\mu$
with a corresponding source we find
\be
0=\sum_x\left\{\nabla^+_0\epsilon_x\left(1+{ia_\tau\over\hbar}
\sigma_{0,x}\right){\delta W_0[\sigma,\tilde\sigma]\over\psi_{0,x}}
+\sum_j\nabla^+_j\epsilon_x
\left[\left(1+{a^2\over\hbar^2}\tilde\sigma_{j,x}\right)
{\delta W_0[\sigma,\tilde\sigma]\over\psi_{j,x}}-{1\over m}\sigma_{j,x}
{\delta W_0[\sigma,\tilde\sigma]\over\delta\tilde\sigma_{j,x}}\right]\right\},
\ee
After integration in part, omitting the boundary terms
and averaging over the disorder we have
\bea
0&=&\int\cd{v}e^{-{1\over2g}\int_\mb{x}v^2_\mb{x}}\Biggl\{
\nabla^-_0\left[\left(1+{ia_\tau\over\hbar}(\sigma_{0,x}-iv_\mb{x})\right)
{\delta W_0[\sigma_0-iv,\sigma_j,\tilde\sigma]\over\psi_{0,x}}\right]
\nonu
&&+\sum_j\nabla^-_j\left[\left(1+{a^2\over\hbar^2}\tilde\sigma_{j,x}\right)
{\delta W_0[\sigma_0-iv,\sigma_j,\tilde\sigma]\over\psi_{j,x}}
-{1\over m}\sigma_{j,x}{\delta W_0[\sigma_0-iv,\sigma_j,\tilde\sigma]
\over\delta\tilde\sigma_{j,x}}\right]\Biggr\}.
\eea
By taking the functional
derivative of the 'equation of motion' for the impurity averaging,
\be
0=\int\cd{v}e^{-{1\over2g}\int_\mb{x}v^2_\mb{x}}\left(i\int_t
{\delta W_0[\sigma_0-iv,\sigma_j,\tilde\sigma]\over\psi_{0,t,\mb{x}}}
+{1\over g}v_\mb{x}W_0[\sigma_0+v,\sigma_j,\tilde\sigma]\right)
\ee
we can write find the Ward identity
\be\label{ward}
0=\nabla^-_0\left[\left(1+{ia_\tau\over\hbar}\sigma_{0,x}\right)
{\delta W[\sigma,\tilde\sigma]\over\psi_{0,x}}
-{ga_\tau\over\hbar}\int_t{\delta^2W[\sigma,\tilde\sigma]
\over\psi_{0,t,\mb{x}}\psi_{0,x}}\right]
+\sum_j\nabla^-_j\left[\left(1+{a^2\over\hbar^2}\tilde\sigma_{j,x}\right)
{\delta W[\sigma,\tilde\sigma]\over\psi_{j,x}}-{1\over m}\sigma_{j,x}
{\delta W[\sigma,\tilde\sigma]\over\delta\tilde\sigma_{j,x}}\right].
\ee
By assuming the form
\be
W[\sigma]=W_0+\sigma\cdot\rho^*-\hf\sigma\cdot\tD\cdot\sigma
+{1\over3!}W_{\tilde\alpha,\tilde\beta,\tilde\gamma}
\sigma_{\tilde\alpha}\sigma_{\tilde\beta}\sigma_{\tilde\gamma}
+\ord{\sigma^4}
\ee
the first functional derivative of \eq{ward} with respect to $\sigma$ taken at
$\sigma=\tilde\sigma=0$ yields the conserved currents
\be
J_{0,x}=\tD^{\kappa,0}_{y,x}
-{ia_\tau\over\hbar}\delta^{\kappa,0}\delta_{y,x}\rho^*
+{ga_\tau\over\hbar}\int_tW_{(0,t,\mb{x}),(0,x),(\kappa,y)},~~~
J_{j,x}=\tD^{\kappa,j}_{y,x}
+{1\over m}\delta^{\kappa,j}\delta_{y,x}\tilde\rho^*_j
\ee
for arbitrary $\kappa$ and $y$. For U.V. finite model or when the
U.V. divergences are logarithmic only the terms $\ord{a_\tau}$
can be neglected.

\section{Functional derivatives}\label{gradexpa}
We give few details of the computation of the second functional
derivative of the effective action \eq{geffact}.
The effective action is supposed to be the sum of terms like
\be
\gamma[\rho]=\int_xV_{A,\mu}(\rho_x)\partial^A\rho_{\mu,x}.
\ee
and the first four functional derivatives are the following
\bea
{\delta\gamma[\rho]\over\delta\rho_{\alpha,a}}
&=&\int_z\left[\delta_{a,z}\partial_{\rho_\alpha} V_{A,\mu}\partial^A\rho_{\mu,z}
+V_{A,\alpha}\partial^A\delta_{a,z}\right]\nonu
{\delta^2\gamma[\rho]\over\delta\rho_{\alpha,a}\delta\rho_{\beta,b}}
&=&\int_z\left[
\delta_{a,z}\delta_{b,z}\partial_{\rho_\beta}\partial_{\rho_\alpha} V_{A,\mu}\partial^A\rho_{\mu,z}
+\delta_{a,z}\partial_{\rho_\alpha} V_{A,\beta}\partial^A\delta_{b,z}
+\delta_{b,z}\partial_{\rho_\beta} V_{A,\alpha}\partial^A\delta_{a,z}\right]\nonu
{\delta^3\gamma[\rho]\over\delta\rho_{\alpha,a}\delta\rho_{\beta,b}\delta\rho_{\gamma,c}}
&=&\int_z\biggl[
\delta_{a,z}\delta_{b,z}\delta_{c,z}\partial_{\rho_\gamma}\partial_{\rho_\beta}\partial_{\rho_\alpha} V_{A,\mu}\partial^A\rho_{\mu,z}
+\delta_{a,z}\delta_{b,z}\partial_{\rho_\beta}\partial_{\rho_\alpha} V_{A,\gamma}\partial^A\delta_{c,z}\nonu
&&+\delta_{a,z}\delta_{c,z}\partial_{\rho_\gamma}\partial_{\rho_\alpha} V_{A,\beta}\partial^A\delta_{b,z}
+\delta_{b,z}\delta_{c,z}\partial_{\rho_\gamma}\partial_{\rho_\beta} V_{A,\alpha}\partial^A\delta_{a,z}\biggr]\nonu
{\delta^4\gamma[\rho]\over\delta\rho_{\alpha,a}\delta\rho_{\beta,b}\delta\rho_{\gamma,c}\delta\rho_{\epsilon,e}}
&=&\int_z\biggl[
\delta_{a,z}\delta_{b,z}\delta_{c,z}\delta_{e,z}\partial_{\rho_\epsilon}\partial_{\rho_\gamma}\partial_{\rho_\beta}\partial_{\rho_\alpha} V_{A,\mu}\partial^A\rho_{\mu,z}
+\delta_{a,z}\delta_{b,z}\delta_{c,z}\partial_{\rho_\gamma}\partial_{\rho_\beta}\partial_{\rho_\alpha} V_{A,\epsilon}\partial^A\delta_{e,z}\nonu
&&+\delta_{a,z}\delta_{b,z}\delta_{e,z}\partial_{\rho_\epsilon}\partial_{\rho_\beta}\partial_{\rho_\alpha} V_{A,\gamma}\partial^A\delta_{c,z}
+\delta_{a,z}\delta_{c,z}\delta_{e,z}\partial_{\rho_\epsilon}\partial_{\rho_\gamma}\partial_{\rho_\alpha} V_{A,\beta}\partial^A\delta_{b,z}\nonu
&&+\delta_{b,z}\delta_{c,z}\delta_{e,z}\partial_{\rho_\epsilon}\partial_{\rho_\gamma}\partial_{\rho_\beta} V_{A,\alpha}\partial^A\delta_{a,z}\biggr]\nonu
\eea
We write $V_{A,\sigma}=\rho_\kappa V_{\kappa,A,\sigma}$ and find
\bea
\partial_{\rho_\alpha} V_{A,\sigma}
&=&V_{\alpha,A,\sigma}+\rho_\kappa\partial_{\rho_\alpha} V_{\kappa,A,\sigma}\nonu
\partial_{\rho_\beta}\partial_{\rho_\alpha} V_{A,\sigma}
&=&\partial_{\rho_\beta} V_{\alpha,A,\sigma}+\partial_{\rho_\alpha} V_{\beta,A,\sigma}+\rho_\kappa\partial_{\rho_\beta}\partial_{\rho_\alpha} V_{\kappa,A,\sigma}\nonu
\partial_{\rho_\gamma}\partial_{\rho_\beta}\partial_{\rho_\alpha} V_{A,\sigma}
&=&\partial_{\rho_\gamma}\partial_{\rho_\beta} V_{\alpha,A,\sigma}+\partial_{\rho_\gamma}\partial_{\rho_\alpha} V_{\beta,A,\sigma}
+\partial_{\rho_\beta}\partial_{\rho_\alpha} V_{\gamma,A,\sigma}+\rho_\kappa\partial_{\rho_\gamma}\partial_{\rho_\beta}\partial_{\rho_\alpha} V_{\kappa,A,\sigma}\nonu
\partial_{\rho_\epsilon}\partial_{\rho_\gamma}\partial_{\rho_\beta}\partial_{\rho_\alpha} V_{A,\sigma}
&=&\partial_{\rho_\epsilon}\partial_{\rho_\gamma}\partial_{\rho_\beta} V_{\alpha,A,\sigma}
+\partial_{\rho_\epsilon}\partial_{\rho_\gamma}\partial_{\rho_\alpha} V_{\beta,A,\sigma}
+\partial_{\rho_\epsilon}\partial_{\rho_\beta}\partial_{\rho_\alpha} V_{\gamma,A,\sigma}\nonu
&&+\partial_{\rho_\gamma}\partial_{\rho_\beta}\partial_{\rho_\alpha} V_{\epsilon,A,\sigma}
+\rho_\kappa\partial_{\rho_\epsilon}\partial_{\rho_\gamma}\partial_{\rho_\beta}
\partial_{\rho_\alpha} V_{\kappa,A,\sigma}
\eea
and
\bea
{\delta\gamma[\rho]\over\delta\rho_{\alpha,a}}
&=&\int_z\left[\delta_{a,z}V_{\alpha,A,\sigma}\partial^A\rho_{\sigma,z}
+\delta_{a,z}\rho_\kappa\partial_{\rho_\alpha} V_{\kappa,A,\sigma}\partial^A\rho_{\sigma,z}
+\rho_\kappa V_{\kappa,A,\alpha}\partial^A\delta_{a,z}\right]\nonu
&=&\int_z\left[\delta_{a,z}V_{\alpha,A,\sigma}\partial^A\rho_{\sigma,z}
+\delta_{a,z}\rho_\kappa\partial_{\rho_\alpha} V_{\kappa,A,\sigma}\partial^A\rho_{\sigma,z}
+(-1)^{|A|}\delta_{a,z}\partial^A(\rho_\kappa V_{\kappa,A,\alpha})\right]\nonu
{\delta^2\gamma[\rho]\over\delta\rho_{\alpha,a}\delta\rho_{\beta,b}}
&=&\int_z\biggl[
\delta_{a,z}\delta_{b,z}(\partial_{\rho_\beta} V_{\alpha,A,\sigma}+\partial_{\rho_\alpha} V_{\beta,A,\sigma}
+\rho_\kappa\partial_{\rho_\beta}\partial_{\rho_\alpha} V_{\kappa,A,\sigma})\partial^A\rho_{\sigma,z}\nonu
&&+\delta_{a,z}(V_{\alpha,A,\beta}+\rho_\kappa\partial_{\rho_\alpha} V_{\kappa,A,\beta})\partial^A\delta_{b,z}
+\delta_{b,z}(V_{\beta,A,\alpha}+\rho_\kappa\partial_{\rho_\beta} V_{\kappa,A,\alpha})\partial^A\delta_{a,z}
\biggr]\nonu
{\delta^3\gamma[\rho]\over\delta\rho_{\alpha,a}\delta\rho_{\beta,b}\delta\rho_{\gamma,c}}
&=&\int_z\biggl[
\delta_{a,z}\delta_{b,z}\delta_{c,z}(\partial_{\rho_\gamma}\partial_{\rho_\beta} V_{\alpha,A,\sigma}+\partial_{\rho_\gamma}\partial_{\rho_\alpha} V_{\beta,A,\sigma}
+\partial_{\rho_\beta}\partial_{\rho_\alpha} V_{\gamma,A,\sigma}+\rho_\kappa\partial_{\rho_\gamma}\partial_{\rho_\beta}\partial_{\rho_\alpha} V_{\kappa,A,\sigma})
\partial^A\rho_{\sigma,z}\nonu
&&+(\delta_{a,z}\delta_{b,z}(\partial_{\rho_\beta} V_{\alpha,A,\gamma}+\partial_{\rho_\alpha} V_{\beta,A,\gamma}+\rho_\kappa\partial_{\rho_\beta}\partial_{\rho_\alpha} V_{\kappa,A,\gamma})
\partial^A\delta_{c,z}+perms.)\biggr]\nonu
{\delta^4\gamma[\rho]\over\delta\rho_{\alpha,a}\delta\rho_{\beta,b}\delta\rho_{\gamma,c}\delta\rho_{\epsilon,e}}
&=&\int_z\biggl[
\delta_{a,z}\delta_{b,z}\delta_{c,z}\delta_{e,z}(\partial_{\rho_\epsilon}\partial_{\rho_\gamma}\partial_{\rho_\beta} V_{\alpha,A,\sigma}+\partial_{\rho_\epsilon}\partial_{\rho_\gamma}\partial_{\rho_\alpha} V_{\beta,A,\sigma}
+\partial_{\rho_\epsilon}\partial_{\rho_\beta}\partial_{\rho_\alpha} V_{\gamma,A,\sigma}+\partial_{\rho_\gamma}\partial_{\rho_\beta}\partial_{\rho_\alpha} V_{\epsilon,A,\sigma}\nonu
&&+\rho_\kappa\partial_{\rho_\epsilon}\partial_{\rho_\gamma}\partial_{\rho_\beta}\partial_{\rho_\alpha} V_{\kappa,A,\sigma})
\partial^A\rho_{\sigma,z}
+(\delta_{a,z}\delta_{b,z}\delta_{c,z}
(\partial_{\rho_\gamma}\partial_{\rho_\beta} V_{\alpha,A,\epsilon}
+\partial_{\rho_\gamma}\partial_{\rho_\alpha} V_{\beta,A,\epsilon}\nonu
&&+\partial_{\rho_\beta}\partial_{\rho_\alpha} V_{\gamma,A,\epsilon}+\rho_\kappa\partial_{\rho_\gamma}\partial_{\rho_\beta}\partial_{\rho_\alpha} V_{\kappa,A,\epsilon})
\partial^A\delta_{e,z}+perms.\biggr],
\eea
where $|A|=|\nu_0|+\cdots$. For homogeneous configurations
$\rho_{\alpha,x}=\rho_\alpha$ we find
\bea
{\delta\gamma[\rho]\over\delta\rho_{\alpha,a}}&=&0\nonu
{\delta^2\gamma[\rho]\over\delta\rho_{\alpha,a}\delta\rho_{\beta,b}}
&=&\int_z\biggl[
\delta_{a,z}(V_{\alpha,A,\beta}+\rho_\kappa\partial_{\rho_\alpha} V_{\kappa,A,\beta})\partial^A\delta_{b,z}
+\delta_{b,z}(V_{\beta,A,\alpha}+\rho_\kappa\partial_{\rho_\beta} V_{\kappa,A,\alpha})\partial^A\delta_{a,z}
\biggr]\nonu
{\delta^3\gamma[\rho]\over\delta\rho_{\alpha,a}\delta\rho_{\beta,b}\delta\rho_{\gamma,c}}
&=&\int_z\biggl[\delta_{a,z}\delta_{b,z}(\partial_{\rho_\beta} V_{\alpha,A,\gamma}
+\partial_{\rho_\alpha} V_{\beta,A,\gamma}
+\rho_\kappa\partial_{\rho_\beta}\partial_{\rho_\alpha} V_{\kappa,A,\gamma})
\partial^A\delta_{c,z}+perms.\biggr]\nonu
{\delta^4\gamma[\rho]\over\delta\rho_{\alpha,a}
\delta\rho_{\beta,b}\delta\rho_{\gamma,c}\delta\rho_{\epsilon,e}}
&=&\int_z\biggl[\delta_{a,z}\delta_{b,z}\delta_{c,z}
(\partial_{\rho_\gamma}\partial_{\rho_\beta} V_{\alpha,A,\epsilon}
+\partial_{\rho_\gamma}\partial_{\rho_\alpha} V_{\beta,A,\epsilon}
+\partial_{\rho_\beta}\partial_{\rho_\alpha} V_{\gamma,A,\epsilon}
+\rho_\kappa\partial_{\rho_\gamma}\partial_{\rho_\beta}
\partial_{\rho_\alpha} V_{\kappa,A,\epsilon})\partial^A\nonu
&&+perms.\biggr]
\eea
or in Fourier space
\bea
{\delta^2\gamma[\rho]\over\delta\rho_{\alpha,p}\delta\rho_{\beta,q}}
&=&\int_{a,b,z}e^{iap+ibq}\delta_{a,z}\delta_{b,z}\left[
(V_{\alpha,A,\beta}+\rho_\kappa\partial_{\rho_\alpha} V_{\kappa,A,\beta})(iq)^A
+(V_{\beta,A,\alpha}+\rho_\kappa\partial_{\rho_\beta} V_{\kappa,A,\alpha})(ip)^A\right]\nonu
&=&\delta_{p+q,0}\left[
(V_{\alpha,A,\beta}+\rho_\kappa\partial_{\rho_\alpha} V_{\kappa,A,\beta})(iq)^A
+(V_{\beta,A,\alpha}+\rho_\kappa\partial_{\rho_\beta} V_{\kappa,A,\alpha})(ip)^A\right]\nonu
{\delta^3\gamma[\rho]\over\delta\rho_{\alpha,p}\delta\rho_{\beta,q}\delta\rho_{\gamma,r}}
&=&\delta_{p+q+r,0}\left[
(\partial_{\rho_\beta} V_{\alpha,A,\gamma}+\partial_{\rho_\alpha} V_{\beta,A,\gamma}+\rho_\kappa\partial_{\rho_\beta}\partial_{\rho_\alpha} V_{\kappa,A,\gamma})
(ir)^A+perms.\right]\nonu
{\delta^4\gamma[\rho]\over\delta\rho_{\alpha,p}\delta\rho_{\beta,q}\delta\rho_{\gamma,r}\delta\rho_{\epsilon,s}}
&=&\delta_{p+q+r+s,0}\biggl[
(\partial_{\rho_\gamma}\partial_{\rho_\beta} V_{\alpha,A,\epsilon}+\partial_{\rho_\gamma}\partial_{\rho_\alpha} V_{\beta,A,\epsilon}
+\partial_{\rho_\beta}\partial_{\rho_\alpha} V_{\gamma,A,\epsilon}+\rho_\kappa\partial_{\rho_\gamma}\partial_{\rho_\beta}\partial_{\rho_\alpha} V_{\kappa,A,\epsilon})
(is)^A\nonu
&&+perms.\biggr].
\eea

The first four functional derivatives of the full effective action
on homogeneous background are given by
\bea
{\delta^2\Gamma[\rho]\over\delta\rho_{\alpha,p}\delta\rho_{\beta,q}}
&=&\delta_{p+q,0}\Gamma^{(0)\alpha,\beta}_{p,q}\nonu
{\delta^3\Gamma[\rho]\over\delta\rho_{\alpha,p}
\delta\rho_{\beta,q}\delta\rho_{\gamma,r}}
&=&\Gamma^{\alpha,\beta,\gamma}_{p,q,r}\delta_{p+q+r,0},\nonu
{\delta^4\Gamma[\rho]\over\delta\rho_{\alpha,p}\delta\rho_{\beta,q}\delta\rho_{\gamma,r}\delta\rho_{\epsilon,s}}
&=&\Gamma^{\alpha,\beta,\gamma,\epsilon}_{p,q,r,s}\delta_{p+q+r+s,0},
\eea
where
\bea
\Gamma^{(0)\alpha,\beta}_{p,q}&=&
V_{\alpha,\beta}(q)+\rho_\kappa\partial_{\rho_\alpha} V_{\kappa,\beta}(q)
+V_{\beta,\alpha}(p)+\rho_\kappa\partial_{\rho_\beta} V_{\kappa,\alpha}(p)
-\partial_{\rho_\alpha}\partial_{\rho_\beta} U\nonu
&=&V_{\alpha,\beta}(q)+V_{\beta,\alpha}(-q)+\rho_\kappa\partial_{\rho_\alpha} V_{\kappa,\beta}(q)
+\rho_\kappa\partial_{\rho_\beta} V_{\kappa,\alpha}(-q)-\partial_{\rho_\alpha}\partial_{\rho_\beta} U\nonu
\Gamma^{\alpha,\beta,\gamma}_{p,q,r}&=&
(\partial_{\rho_\beta} V_{\alpha,\gamma}(r)+\partial_{\rho_\alpha} V_{\beta,\gamma}(r)
+\rho_\kappa\partial_{\rho_\beta}\partial_{\rho_\alpha} V_{\kappa,\gamma}(r)
+perms.)-\partial_{\rho_\alpha}\partial_{\rho_\beta}\partial_{\rho_\gamma} U,\nonu
&=&\rho_\kappa\partial_{\rho_\beta}\partial_{\rho_\alpha} V_{\kappa,\gamma}(r)
+\partial_{\rho_\beta} V_{\alpha,\gamma}(r)+\partial_{\rho_\alpha} V_{\beta,\gamma}(r)
+\rho_\kappa\partial_{\rho_\alpha}\partial_{\rho_\gamma} V_{\kappa,\beta}(q)
+\partial_{\rho_\gamma} V_{\alpha,\beta}(q)+\partial_{\rho_\alpha} V_{\gamma,\beta}(q)\nonu
&&+\rho_\kappa\partial_{\rho_\beta}\partial_{\rho_\gamma} V_{\kappa,\alpha}(p)
+\partial_{\rho_\gamma} V_{\beta,\alpha}(p)+\partial_{\rho_\beta} V_{\gamma,\alpha}(p)
-\partial_{\rho_\alpha}\partial_{\rho_\beta}\partial_{\rho_\gamma} U,\nonu
\Gamma^{\alpha,\beta,\gamma,\epsilon}_{p,q,r,s}&=&
\partial_{\rho_\gamma}\partial_{\rho_\beta} V_{\alpha,\epsilon}(s)
+\partial_{\rho_\gamma}\partial_{\rho_\alpha} V_{\beta,\epsilon}(s)
+\partial_{\rho_\beta}\partial_{\rho_\alpha} V_{\gamma,\epsilon}(s)
+\rho_\kappa\partial_{\rho_\gamma}\partial_{\rho_\beta}\partial_{\rho_\alpha} V_{\kappa,\epsilon}(s)
+perms.-\partial_{\rho_\alpha}\partial_{\rho_\beta}\partial_{\rho_\gamma}\partial_{\rho_\epsilon} U\nonu
&=&\rho_\kappa\partial_{\rho_\beta}\partial_{\rho_\alpha}\partial_{\rho_\gamma} V_{\kappa,\epsilon}(s)
+\partial_{\rho_\alpha}\partial_{\rho_\gamma} V_{\beta,\epsilon}(s)
+\partial_{\rho_\beta}\partial_{\rho_\gamma} V_{\alpha,\epsilon}(s)
+\partial_{\rho_\beta}\partial_{\rho_\alpha} V_{\gamma,\epsilon}(s)\nonu
&&+(\rho_\kappa\partial_{\rho_\beta}\partial_{\rho_\alpha}\partial_{\rho_\epsilon} V_{\kappa,\gamma}(r)
+\partial_{\rho_\alpha}\partial_{\rho_\epsilon} V_{\beta,\gamma}(r)
+\partial_{\rho_\beta}\partial_{\rho_\epsilon} V_{\alpha,\gamma}(r)
+\partial_{\rho_\beta}\partial_{\rho_\alpha} V_{\epsilon,\gamma}(r)\nonu
&&+(\rho_\kappa\partial_{\rho_\beta}\partial_{\rho_\gamma}\partial_{\rho_\epsilon} V_{\kappa,\alpha}(p)
+\partial_{\rho_\gamma}\partial_{\rho_\epsilon} V_{\beta,\alpha}(p)
+\partial_{\rho_\beta}\partial_{\rho_\epsilon} V_{\gamma,\alpha}(p)
+\partial_{\rho_\beta}\partial_{\rho_\gamma} V_{\epsilon,\alpha}(p)\nonu
&&+(\rho_\kappa\partial_{\rho_\alpha}\partial_{\rho_\gamma}\partial_{\rho_\epsilon} V_{\kappa,\beta}(q)
+\partial_{\rho_\gamma}\partial_{\rho_\epsilon} V_{\alpha,\beta}(q)
+\partial_{\rho_\alpha}\partial_{\rho_\epsilon} V_{\gamma,\beta}(q)
+\partial_{\rho_\alpha}\partial_{\rho_\gamma} V_{\epsilon,\beta}(q)
-\partial_{\rho_\alpha}\partial_{\rho_\beta}\partial_{\rho_\gamma}\partial_{\rho_\epsilon} U.
\eea
and the dependence on $\rho$ is suppresed. Therefore the second
functional derivative up to $\ord{\delta\rho^2}$ is
\bea
{\delta^2\Gamma[\rho]\over\delta\rho_{\alpha,p}\delta\rho_{\beta,q}}
&=&\delta_{p+q,0}\Gamma^{(0)\alpha,\beta}_{p,q}+\Gamma^{(1)\alpha,\beta}_{p,q}
+\hf\Gamma^{(2)\alpha,\beta}_{p,q}+\ord{\delta\rho^3}\nonu
\Gamma^{(1)\alpha,\beta}_{p,q}&=&\int_r\delta_{p+q+r,0}
\delta\rho_{\gamma,r}\Gamma_{p,q,r}^{\alpha,\beta,\gamma},\nonu
\Gamma^{(2)\alpha,\beta}_{p,q}&=&\int_{r,s}\delta_{p+q+r+s,0}
\delta\rho_{\gamma,r}\delta\rho_{\rho,s}\Gamma^{\alpha,\beta,\gamma,\rho}_{p,q,r,s}.
\eea

The appplication of these formulae for the effective action
\eq{noncov} yields the second derivative
\be
\Gamma^{(0)\alpha,\beta}_{p,q}=\pmatrix{\Gamma^t
+\hf\nabla^0_1\Gamma^{tt}(p)+\hf\nabla^0_1\Gamma^{tt}(q)&
i\mb{p}\hf\Gamma^{st}(p)+i\mb{q}\hf\nabla^0_1\Gamma^{ts}(q)\cr
i\mb{q}\hf\Gamma^{st}(q)+i\mb{p}\hf\nabla^0_1\Gamma^{ts}(p)&
\hf\Gamma^T(q)T_q+\hf\Gamma^L(q)L_q+\hf\Gamma^T(p)T_p+\hf\Gamma^L(p)L_p}
\ee
where $\Gamma^t=-\partial_0^2U$, $\Gamma^s=-\partial_1^2U$. The propagator is
\be
\delta_{p+q,0}\Gamma^{(0)-1\alpha,\beta}_{p,q}
=\pmatrix{{\Gamma^L\over\Gamma_D}&-i\mb{q}\hf{\nabla^0_2\Gamma^{ts}\over\Gamma_D}\cr
-i\mb{q}\hf{\nabla^0_2\Gamma^{ts}\over\Gamma_D}&{1\over\Gamma^T}T
+{\Gamma^t+\nabla^0_1\Gamma^{tt}\over\Gamma_D}L}
\ee
with
\be
\Gamma_D=(\Gamma^t+\nabla^0_1\Gamma^{tt})\Gamma^L
+{\mb{q}^2\over4}(\nabla^0_2\Gamma^{ts})^2.
\ee

The third and fourth derivatives are of the form
\bea
\Gamma_{p,q,r}^{0,0,0}&=&\nabla^1_2V_{0,0}(p)+\nabla^1_2V_{0,0}(q)
+\nabla^1_2V_{0,0}(r)+\partial_0\Gamma^t\nonu
&=&\hf\nabla^1_2\Gamma^{tt}(p)+\hf\nabla^1_2\Gamma^{tt}(q)
+\hf\nabla^1_2\Gamma^{tt}(r)+\partial_0\Gamma^t\nonu
\Gamma_{p,q,r}^{0,0,\ell}&=&\nabla^1_2V_{0,\ell}(r)
+\partial_0V_{\ell,0}(p)+\partial_0V_{\ell,0}(q)+\partial_\ell\Gamma^t\nonu
&=&{i\over2}\nabla^1_2\Gamma^{ts}(r)r_\ell+{i\over2}\partial_0\Gamma^{st}(p)p_\ell
+{i\over2}\partial_0\Gamma^{st}(q)q_\ell+\partial_\ell\Gamma^t\nonu
\Gamma_{p,q,r}^{0,k,\ell}&=&\partial_0V_{k,\ell}(r)+\partial_0V_{\ell,k}(q)
+\delta_{k,\ell}\partial_0\Gamma^s\nonu
&=&\hf\partial_0\Gamma^T(r)T_r^{k,\ell}+\hf\partial_0\Gamma^T(q)T_q^{\ell,k}
+\hf\partial_0\Gamma^L(r)L_r^{k,\ell}+\hf\partial_0\Gamma^L(q)L_q^{\ell,k}
+\delta_{k,\ell}\partial_0\Gamma^s\nonu
\Gamma^{0,0,0,0}_{p,q,r,s}&=&\nabla^2_3V_{0,0}(p)+\nabla^2_3V_{0,0}(q)
+\nabla^2_3V_{0,0}(r)+\nabla^2_3V_{0,0}(s)+\partial^2_0\Gamma^t\nonu
&=&\hf\nabla^2_3\Gamma^{tt}(p)+\hf\nabla^2_3\Gamma^{tt}(q)
+\hf\nabla^2_3\Gamma^{tt}(r)+\hf\nabla^2_3\Gamma^{tt}(s)+\partial^2_0\Gamma^t\nonu
\Gamma^{0,0,0,k}_{p,q,r,s}&=&\nabla^2_3V_{0,k}(s)+\partial^2_0V_{k,0}(p)
+\partial^2_0V_{k,0}(q)+\partial^2_0V_{k,0}(r)+\partial_0\partial_k\Gamma^t\nonu
&=&{i\over2}\nabla^2_3\Gamma^{ts}(s)s^k+{i\over2}\partial^2_0\Gamma^{st}(p)p^k
+{i\over2}\partial^2_0\Gamma^{st}(q)q^k+{i\over2}\partial^2_0\Gamma^{st}(r)r^k
+\partial_0\partial_k\Gamma^t\nonu
\Gamma^{0,0,k,\ell}_{p,q,r,s}&=&\partial^2_0V_{k,\ell}(s)
+\partial^2_0V_{\ell,k}(r)+\delta^{k,\ell}\partial^2_0\Gamma^s\nonu
&=&\hf\partial^2_0\Gamma^T(s)T_s^{k,\ell}+\hf\partial^2_0\Gamma^L(s)L_s^{k,\ell}
+\hf\partial^2_0\Gamma^T(r)T_r^{k,\ell}+\hf\partial^2_0\Gamma^L(r)L_r^{k,\ell}
+\delta^{k,\ell}\partial^2_0\Gamma^s.
\eea

\section{One-loop $W[\sigma]$}\label{pertexp}
This Appendix we briefly record the expressions needed for
the computation of the Green functions occuring in the one-loop
approximation of the generator funtional $W[\sigma]$. This generator
functional is non-trivial even for free electrons due
to the exchange interaction.

\subsection{Densities}
We start with the expresions for the densities
\be
\rho^*=-{i\hbar\over2}G_{t,\mb{x},t+\delta,\mb{x}}+c.c.
=\int_\mb{q}\Theta(-E_\mb{q})
\ee
and
\be
\tilde\rho^*={1\over3}\sum_j\tilde\rho^*_j
=-{i\hbar\over6}\sum_{j>0}G_{t,\mb{x},t+\delta,\mb{x}+j}+c.c.
=\int_\mb{q}\Theta(-E_\mb{q})\cos q_1a
\ee
in terms of the electron propagator $G_p$. The local potential is given by
\be
{\hbar\over Vi}\tr\log D^{-1}
=-i\int_{\mb{q},E}\log[E-E_\mb{q}+i\epsilon_\mb{q}].
\ee

Charge conjugation acts as $\mu\to\mu^c=6\hbar^2/ma^2-\mu$,
$\rho^*\to\rho^{c*}={\cal B}_\rho-\rho^*$ and $\tilde\rho^*\to\tilde\rho^{c*}=\tilde\rho^*$
where the band contribution is
\be
{\cal B}_\rho=\int_{\mb{q}}1
\ee

\subsection{Two-point functions}
We need
\bea
\tD^{0,0}_p&=&-i\hbar\int_rG_rG_{p+r},\nonu
\tD^{0,j}_p&=&-{i\hbar^2\over ma}e^{-i{p_ja\over2}}\int_r
G_rG_{p+r}\sin a\left(r_j+{p_j\over2}\right)\nonu
\tD^{j,k}_p&=&-{i\hbar^3\over m^2a^2}\int_rG_rG_{p+r}
\sin a\left(r_j+{p_j\over2}\right)\sin a\left(r_k+{p_k\over2}\right),
\eea
where the Fourier transform is defined as
\be
\tD^{\alpha,\beta}_{x,y}=\int_p\tD^{\alpha,\beta}_pe^{ip(x-y)}.
\ee

It is straightforward to perform the energy integrals.
The results are shown for the external momentum
$\hat\mb{p}=(0,0,\hat p)$ after having carried out the
replacement $\mu\to\mu-\sigma_0$,
\bea
\tD^{0,0}(\omega,\mb{p}^2)&=&2\pp\int_\mb{q}
{\Theta(-E_\mb{q})\Delta E_{\mb{q},\mb{p}}\over
(\Delta E_{\mb{q},\mb{p}})^2-\hbar^2\omega^2}
+i\pi\int_\mb{q}\Theta(-E_\mb{q})\Theta(E_{\mb{q}+\mb{p}})
[\delta(\Delta E_{\mb{q},\mb{p}}-\hbar\omega)
+\delta(\Delta E_{\mb{q},\mb{p}}+\hbar\omega)]\nonu
\tD^{0,j}(\omega,\mb{p})&=&-\delta^{j,3}{2\omega\hbar\over am}\pp
\int_\mb{q}{\Theta(-E_\mb{q})\sin a\left(q_3+{p\over2}\right)
\over(\Delta E_{\mb{q},\mb{p}})^2-\hbar^2\omega^2}\nonu
&&-{i\hbar\pi\over am}\int_\mb{q}\Theta(-E_\mb{q})
\Theta(E_{\mb{q}+\mb{p}})\sin a\left(q_j+{p_j\over2}\right)
[\delta(\Delta E_{\mb{q},\mb{p}}-\hbar\omega)-
\delta(\Delta E_{\mb{q},\mb{p}}+\hbar\omega)]\nonu
&=&\tD^{0,j}(-\omega,-\mb{p})=-\tD^{0,j}(-\omega,\mb{p})=\tD^{j,0}(\omega,\mb{p})\nonu
\tD^T(\omega,\mb{p}^2)&=&{2\hbar^2\over a^2m^2}\pp\int_\mb{q}
{\Theta(-E_\mb{q})\sin^2aq_1\Delta E_{\mb{q},\mb{p}}\over
(\Delta E_{\mb{q},\mb{p}})^2-\hbar^2\omega^2}\nonu
&&+{i\pi\hbar^2\over a^2m^2}\int_\mb{q}\Theta(-E_\mb{q})\Theta(E_{\mb{q}+\mb{p}})\sin^2aq_1
[\delta(\Delta E_{\mb{q},\mb{p}}-\hbar\omega)+\delta(\Delta E_{\mb{q},\mb{p}}+\hbar\omega)]\nonu
\tD^L(\omega,\mb{p}^2)&=&{2\hbar^2\over a^2m^2}\pp\int_\mb{q}
{\Theta(-E_\mb{q})\sin^2a\left(q_3+{p\over2}\right)
\Delta E_{\mb{q},\mb{p}}\over(\Delta E_{\mb{q},\mb{p}})^2-\hbar^2\omega^2}\nonu
&&+{i\pi\hbar^2\over a^2m^2}\int_\mb{q}\Theta(-E_\mb{q})\Theta(E_{\mb{q}+\mb{p}})
\sin^2a\left(q_1+{p\over2}\right)
[\delta(\Delta E_{\mb{q},\mb{p}}-\hbar\omega)
+\delta(\Delta E_{\mb{q},\mb{p}}+\hbar\omega)],
\eea
where $\Delta E_{\mb{q},\mb{p}}=E_\mb{q}-E_{\mb{q},\mb{p}}$ and
$\pp$ denotes the principal value prescription. The singularity and
the Dirac delta in the integrand render these expressions
unpractical for numerical evaluation. We used Wick rotated
integrals and considered the dependence in Euclidean energies
in the numerical work. The corresponding expressions are
\bea\label{initialc}
\tD^{0,0}(\omega,\mb{p}^2)&=&2\int_\mb{q}{\Theta(-E_\mb{q})
\Delta E_{\mb{q},\mb{p}}\over(\Delta E_{\mb{q},\mb{p}})^2+\hbar^2\omega^2}\nonu
\tD^{0,j}(\omega,\mb{p})&=&-{2i\omega\hbar^2\over am}
\int_\mb{q}{\Theta(-E_\mb{q})\sin a\left(q_3+{p\over2}\right)
1\over(\Delta E_{\mb{q},\mb{p}})^2+\hbar^2\omega^2}\nonu
\tD^T(\omega,\mb{p}^2)&=&{2\hbar^2\over a^2m^2}
\int_\mb{q}{\Theta(-E_\mb{q})\sin^2aq_1
\Delta E_{\mb{q},\mb{p}}\over(\Delta E_{\mb{q},\mb{p}})^2+\hbar^2\omega^2}\nonu
\tD^L(\omega,\mb{p}^2)&=&{2\hbar^2\over a^2m^2}
\int_\mb{q}{\Theta(-E_\mb{q})\sin^2a\left(q_3+{p\over2}\right)
\Delta E_{\mb{q},\mb{p}}\over(\Delta E_{\mb{q},\mb{p}})^2+\hbar^2\omega^2}.
\eea

These expressions show clearly the need of keeping a non-vanishing
energy $\hbar\omega$ in any computation carried out in a finite system.
In fact, the limit $\omega\to0$ generates a Dirac delta in the integrands
which yields singular results in a finite system where the
density of states is the sum of Dirac delta peaks.

\end{document}